\documentclass{emulateapj}
\usepackage{subfigure}
\usepackage{amsmath} %pacchetto matematico
\usepackage{color}
\usepackage{graphicx}
\usepackage{textcomp}
\usepackage{natbib}
\usepackage{longtable}
\usepackage{footnote}
\pdfoutput=1
\usepackage{ifpdf}

\begin{document}
\title{The connection between the radio jet and the gamma-ray emission in the radio~galaxy 3C~120}
\author{Carolina Casadio\altaffilmark{1}, Jos\'e L. G\'omez\altaffilmark{1}, Paola Grandi\altaffilmark{2}, 
Svetlana G. Jorstad\altaffilmark{3,4}, Alan P. Marscher\altaffilmark{3}, Matthew L. Lister\altaffilmark{5}, 
Yuri Y. Kovalev\altaffilmark{6,7}, Tuomas Savolainen\altaffilmark{7,8}, Alexander B. Pushkarev\altaffilmark{10,9,7} 
}

\altaffiltext{1}{Instituto de Astrof\'{\i}sica de Andaluc\'{\i}a, CSIC, Apartado 3004, 18080,
Granada, Spain}
\altaffiltext{2}{Istituto Nazionale di Astrofisica-IASFBO, Via Gobetti 101, I-40129, Bologna, Italy}
\altaffiltext{3}{Institute for Astrophysical Research, Boston University, 725 Commonwealth Avenue, Boston, MA 02215}
%\email{casadio@iaa.es}
\altaffiltext{4}{Astronomical Institute, St. Petersburg State University, Universitetskij Pr. 28, Petrodvorets, 198504 St. Petersburg, Russia}
\altaffiltext{5}{Department of Physics, Purdue University, 525 Northwestern Avenue, West Lafayette, IN 47907, USA}
\altaffiltext{6}{Astro Space Center, Lebedev Physical Institute, Russian Academy of Sciences, Profsoyuznaya str. 84/32, Moscow 117997, Russia}
\altaffiltext{7}{Max-Planck-Institut f\"ur Radioastronomie, Auf dem H\"ugel 69, 53121 Bonn, Germany}
\altaffiltext{8}{Aalto University Mets\"ahovi Radio Observatory, Mets\"ahovintie 114, 02540 Kylm\"al\"a, Finland}
\altaffiltext{9}{Pulkovo Observatory, Pulkovskoe Chaussee 65/1, 196140 St. Petersburg, Russia}
\altaffiltext{10}{Crimean Astrophysical Observatory, 98409 Nauchny, Crimea, Russia}

\shorttitle{}
\shortauthors{}
\begin{abstract}

  We present the analysis of the radio jet evolution of the radio~galaxy 3C~120 during a period of prolonged $\gamma$-ray activity detected by the {\it Fermi} satellite between December 2012 and October 2014. We find a clear connection between the $\gamma$-ray and radio emission, such that every period of $\gamma$-ray activity is accompanied by the flaring of the mm-VLBI core and subsequent ejection of a new superluminal component. However, not all ejections of components are associated with $\gamma$-ray events detectable by {\it Fermi}. Clear $\gamma-ray$ detections are obtained only when components are moving in a direction closer to our line of sight. This suggests that the observed $\gamma$-ray emission depends not only on the interaction of moving components with the mm-VLBI core, but also on their orientation with respect to the observer. Timing of the $\gamma$-ray detections and ejection of superluminal components locate the $\gamma$-ray production to within $\sim$0.13 pc from the mm-VLBI core, which was previously estimated to lie about 0.24 pc from the central black hole. This corresponds to about twice the estimated extension of the broad line region, limiting the external photon field and therefore suggesting synchrotron self Compton as the most probable mechanism for the production of the $\gamma$-ray emission. Alternatively, the interaction of components with the jet sheath can provide the necessary photon field to produced the observed $\gamma$-rays by Compton scattering.

\end{abstract}

\section{Introduction}

In the unified scheme of active galactic nuclei (AGN), Fanaroff-Riley radio~galaxies of type I (FRI) and II (FRII) are considered the parent population of BL Lacs and flat spectrum radio quasars (FSRQ), respectively. FRI and FRII radio~galaxies belong to the misaligned AGN class, as they are oriented at larger viewing angles than blazars (the most luminous and variable BL Lac objects and flat-spectrum radio quasars). Relativistic beaming amplifies the emission of jets pointing toward the observer, making blazars the brightest objects in the extragalactic sky also in the $\gamma$-ray band. In fact, among the more than one thousand extragalactic sources detected by the Large Area Telescope (LAT) onboard the {\it Fermi} Gamma-Ray Space Telescope in two years, only 3\% of them are not associated to blazar objects. FRI and FRII radio~galaxies fall inside this 3\%, with the predominance of nearby radio~galaxies of type I \citep{Grandi:2012kx}. FRII radio~galaxies with GeV emission are rare, being 3C~111 the only FRII so far with a confirmed {\it Fermi} counterpart. 3C~111 is also the first FRII radio~galaxy where a $\gamma$-ray flare has been associated with the ejection of a new bright knot from the radio core \citep{Grandi:2012fk}. The simultaneity of the observed flare in millimeter, optical, X-ray and $\gamma$-ray bands lead \cite{Grandi:2012fk} to claim that the GeV dissipation region is located at a distance of about 0.3 pc from the central black hole.   

The rapid $\gamma$-ray variability observed in FRI radio~galaxies, with time scales of months in the case of NGC~1275 \citep{Abdo:2010vn, Kataoka:2010zr}, or even days for the case of the TeV variability in M87 \citep{Aharonian:2006fk, Harris:2011uq}, suggests that the $\gamma$-ray emission in these sources originates also in a very compact jet region. Results from over ten years of multiwavelength observations of M87 \citep{Abramowski:2012oq} suggest that the very high energy flares may take place in the core (for the case of the 2008 and 2011 TeV flares), or in the HST-1 complex, about 0.8$^{\prime\prime}$ downstream the jet (in the case of the 2005 TeV flare). More recently, \cite{Hada:2014ys} found that the last TeV flare occurred in M87 in 2012 originates in the jet base, within 0.03 pc from the black hole, but no correlation with the MeV-GeV light curve obtained by {\it Fermi} has been found.

Recently, the Large Area Telescope detected an unprecedented $\gamma$-ray flare from the FRI radio~galaxy 3C~120 on September 24th, 2014 when the source reached a daily flux (E$>$100 MeV) of 1.0$\pm$0.3$\times$10$^{-6}$ photon cm$^{-2}$ s$^{-1}$, as reported in \cite{Tanaka:2014kx}. This recent flare appears to be associated with a higher state of $\gamma$-ray activity in the source. In fact, since December 2012 the radio~galaxy 3C~120 registered a series of $\gamma$-ray events, indicating a flaring activity that lasted until at least our last data analyzed, in 2014 October 4.

The radio~galaxy 3C~120 ($z$=0.033) presents an FRI morphology, but it also has a blazar-like radio jet, showing multiple superluminal components at parsec scales \citep{Gomez:2000ys, Gomez:2001fk, Gomez:2011vn}, as well as at distances as large as 150 pc in projection from the core \citep{Walker:2001uq}. This radio~galaxy reveals also X-ray properties similar to Seyfert galaxies, with the X-ray spectral slope increasing with intensity \citep{Maraschi:1991kx}, and a prominent iron emission line at a photon energy of 6.4 KeV. This implies that most of the X-ray emission comes from, or near the accretion disk, rather than in the jet. In addition, the observed strong correlation between dips in the X-ray emission and the ejection of new superluminal components in the radio jet \citep{Marscher:2002kx, Chatterjee:2009ve} reveals a clear connection between the accretion disk and the radio jet. 

In this paper we present the first association of $\gamma$-ray emission and the ejection of new superluminal components in a FRI radio~galaxy, 3C~120. This resembles the recent findings on the FRII radio galaxy 3C~111 \citep{Grandi:2012fk}. We present the radio data set analyzed in this study, as well as methods used to reduce radio data in Section 2; in Section 3 we present the analysis and results of the {\it Fermi}-LAT data; in Section 4 we study the radio emission at 15 and 43 GHz in the parsec scale jet; Section 5 presents the connection between the $\gamma$-ray and radio emission, and in Section 6 we discuss our findings.

The cosmological values adopted from the Planck's results \citep{Planck-Collaboration:2014} are $\Omega_{m}$= 0.3, $\Omega_{\Lambda}$= 0.7, and ${\it H}_{0}$ = 68 km s$^{-1}$Mpc$^{-1}$. With these values, at the redshift of 3C~120 ($z$=0.033) 1 mas corresponds to a linear distance 0.67 pc, and a proper motion of 1 mas yr$^{-1}$ corresponds to an apparent speed of 2.21$c$.

\section{Radio data analysis}

To study the structure of the radio jet in 3C~120 we have collected data from two of the most extended VLBA monitoring programs: the MOJAVE\footnote{http://www.physics.purdue.edu/MOJAVE/} and the VLBA-BU-BLAZAR programs\footnote{http://www.bu.edu/blazars/research.html}. This radio dataset consists of 46 epochs of VLBA data at 15 GHz taken from the MOJAVE survey, covering the observing period from June 2008 to August 2013, and 21 epochs of VLBA data at 43 GHz from the VLBA-BU-BLAZAR program, covering the period from January 2012 to May 2014. 

The reduction of the VLBA 43 GHz data has been performed using a combination of {\tt AIPS} and {\tt Difmap} packages, as described in \cite{Jorstad:2005fk}. VLBA data at 15 GHz have been calibrated by the MOJAVE team, following the procedure described in \cite{Lister:2009fk}. For comparison across epochs all the images have been convolved with a mean beam of 0.3$\times$0.15 mas and 1.2$\times$0.5 mas for the VLBA-BU-BLAZAR and MOJAVE programs, respectively.
 
  To determine the structural changes in the radio jet we have modeled the radio emission through fitting of the visibilities to circular Gaussian components using {\tt Difmap} \citep{Shepherd:1997fk}. Fitted values for each   component are the flux density, separation and position angle from the core, and size. 
  These are tabulated in two tables at the end of the paper - Tables~\ref{tab15G} and \ref{tab43G} for the 15 and 43 GHz data, respectively.
  %These are tabulated in two distinct tables for the 15 and 43 GHz data, available in the published version of the 
  %paper. 
  %at the end of the paper - Tables~\ref{tab15G} and \ref{tab43G} for the 15 and 43 GHz data, respectively.

 \section {Fermi-LAT data analysis}

The LAT data collected during 72 months of operation (from 2008 August 4 to 2014 August 4)\footnote{Mission Elapsed Time (MET) Start Time=239557417:MET End Time = 428803203} were analyzed using the {\it Fermi}-LAT ScienceTools software (version v9r32p5) and the P7REP$\_$SOURCE$\_$V15 set of instrument response functions \citep{Ackermann:2012fk}\footnote{Science Tools and instrument response functions are available from the Fermi Science Support Center: \ttfamily{http://fermi.gsfc.nasa.gov/ssc/data/analysis}.}. The time intervals  when the rocking angle of the LAT was greater than $52^{\circ}$ were rejected and a cut to select a maximum zenith angle of 100$\degr$ of the events was applied to exclude $\gamma$-rays originating from cosmic ray interactions with the Earth's atmosphere.

The detection significance of a source is provided by  the TS = 2[logL(source)-logL(no~source)], where L(source) is the maximum likelihood value for a model with an additional source at a specified location and L(no~source) is the maximum likelihood value for a model without the additional source\citep{Mattox:1996uq}. When the TS is less than 10 or the ratio of the flux uncertainty to the flux is more than 0.5, a  2$\sigma$ upper limit of the flux is provided. Depending on the TS value, the upper limits are calculated using the profile (TS$\ge1$) or the  Bayesian (TS$<1$) method as  described in the second LAT catalog \citep[2FGL catalog;][]{Nolan:2012kx}. All errors reported in the figures or quoted in the text are 1$\sigma$ statistical errors. The estimated systematic errors on the flux, 10$\%$ at 100 MeV, decreasing to 5$\%$ at 560 MeV, and increasing to 10$\%$ at 10 GeV, refer to uncertainties on the effective area of the instrument\footnote{http://fermi.gsfc.nasa.gov/ssc/data/analysis/LAT$\_$caveats$\_$pass7.html.}.

\begin{figure*}
\centering
\includegraphics[width=0.49\textwidth]{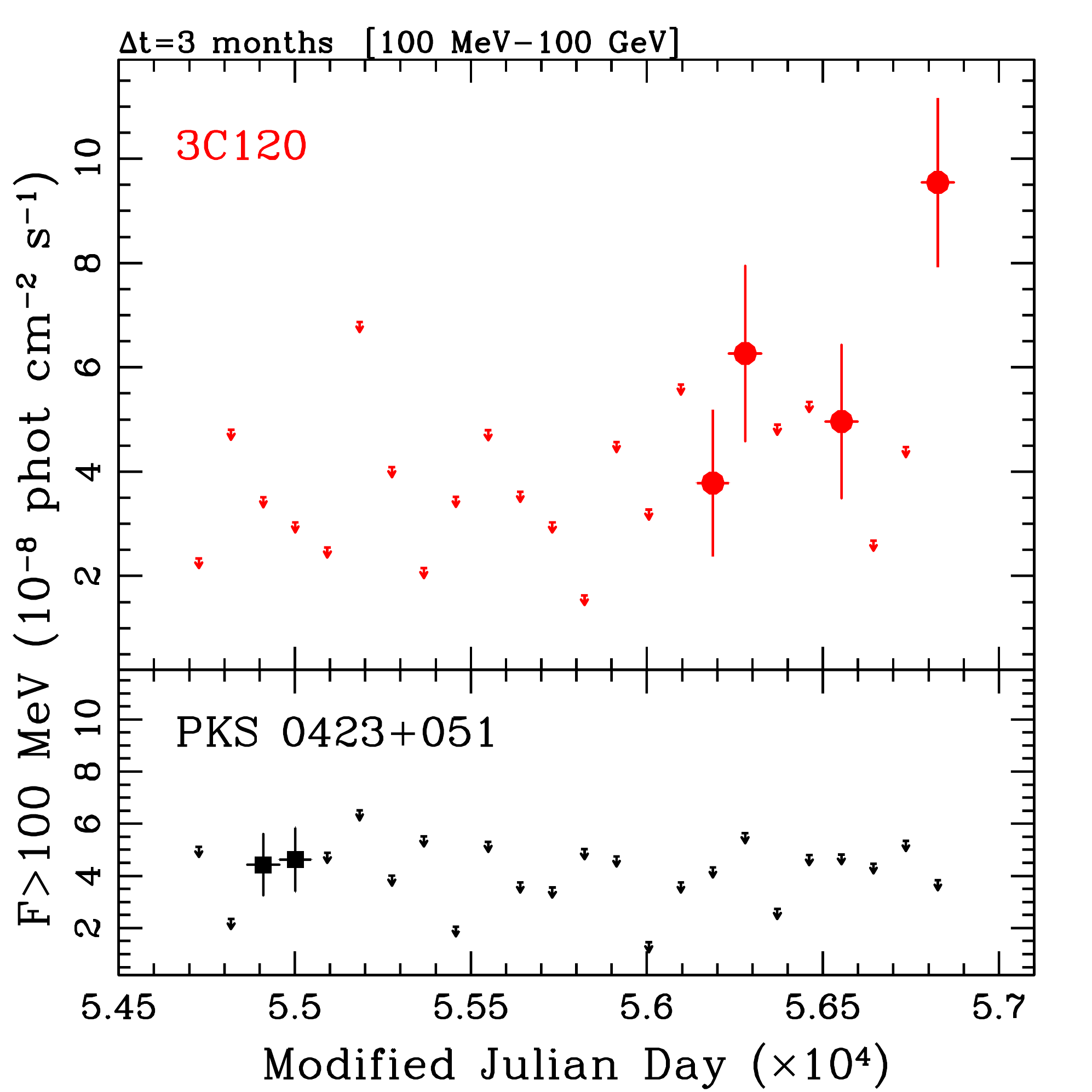}
\includegraphics[width=0.49\textwidth]{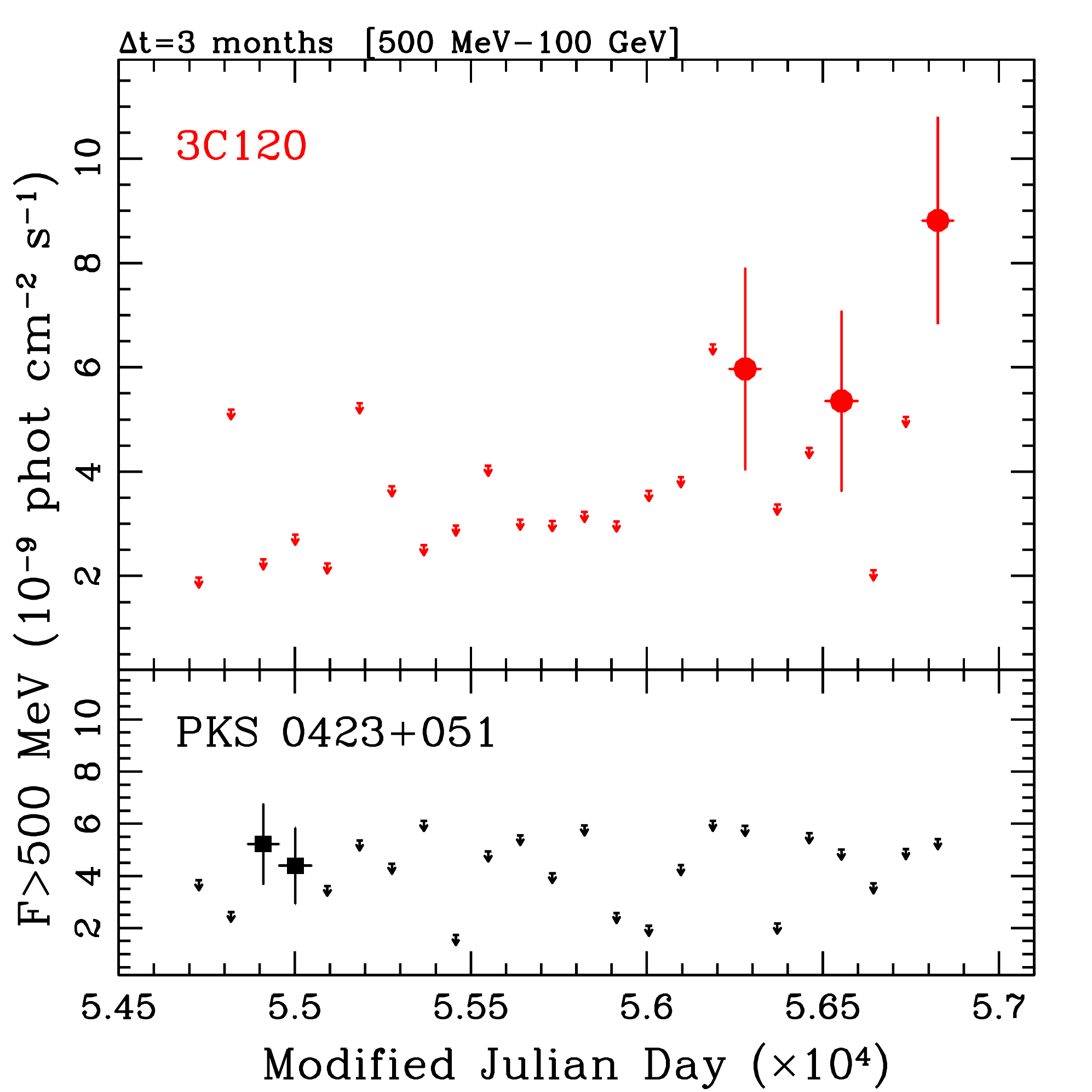}
\caption{{\it Fermi}-LAT light curves of 3C~120 covering 72-month of survey (from 2008 August 4 to 2014 August 6) obtained considering two different energy bands: 100~MeV-100~GeV ({\it left panel}) and 500~MeV-100~GeV ({\it right panel}). A bin width of 3 months is adopted. A 2$\sigma$ upper limit flux is shown (arrow) when the source is not detected (TS$<$10). For comparison the light curves of the nearby FSRQ (PKS~0423$+$051) at 1.6 degrees from 3C~120 are also shown in the same energy ranges. There is not significant overlapping, as 3C~120 and PKS~0423$+$051 appear to be active in different time intervals.}
\label{fermi3m}
\end{figure*}

%\begin{figure*}
%\centering
%\includegraphics[width=0.52\textwidth]{lc15d500.pdf}
%\caption{{\it Fermi}-LAT light curves of 3C~120 and PKS~0423$+$051 covering the time interval (2012-2014) of major activity of 3C~120. The bin size is 15 days and  only photons with E$>500$~MeV are considered. In spite of the short integration time of each bin, 3C~120 is detected six times, between the beginning of 2013 and September - October 2014.}
%\label{fermi15d}
%\end{figure*}

\subsection{The 72-month Average Spectrum}

We performed both binned and unbinned likelihood analysis following the standard LAT data analysis procedures, obtaining consistent results. Here we present results from binned analysis to be consistent with published data~\citep{Ackermann:2011uq}. We accumulated 72 months of data to obtain the average $\gamma$-ray spectral properties of the source. The adopted model included all 2FGL sources within 15$\degr$ of 3C~120 (R.A.(J2000)=68.2962313$\degr$, Dec(J2000)=5.3543389$\degr$).  The studied source was modeled with a power law. All spectral parameters of the sources more than 10$\degr$ from the center of the Region of Interest (RoI) were fixed to the 2FGL values. The Galactic diffuse emission was modeled using the standard diffuse emission model  {\it gll$\_$iem$\_$v05$\_$rev1.fit}  while isotropic $\gamma$-ray emission and the residual cosmic ray contamination in the instrument were modeled using the template  {\it iso$\_$source$\_$v05.txt}.  

  3C~120 was detected in the 100~MeV-100~GeV band with a TS value of 107  ($\sim10 \sigma$). The source is steep ( photon index $\Gamma$=2.7$\pm$0.1) and weak  with a flux of F$_{>100~\mathrm{MeV}} =2.5\pm0.4\times10^{-8}$ photon cm$^{-2}$ s$^{-1}$, in agreement with the result of \cite{Abdo:2010vn}. As noted by the same authors, a Flat Spectrum Quasar,  PKS~0423$+$05, is located at only $\sim1.6$ degrees from the radio galaxy. In our analysis, the blazar is soft  ($\Gamma$=2.6$\pm$0.1) and slightly fainter (F$_{>100~\mathrm{MeV}} =1.77\pm0.36)\times10^{-8}$ photon cm$^{-2}$ s$^{-1}$) than 3C~120.

 \subsection{Light Curves}

Gamma-ray light curves were produced by dividing the analyzed time interval in temporal segments and repeating the likelihood analysis with only the normalizations of the sources within 10$\degr$ free to vary. The spectral slopes of all the sources in the RoI were kept fixed to the best fit values of the 72-month likelihood analysis.  

At first we produced a light curve from 2008 August 4 to 2014 August 4 with a bin size of 3 months in the 100 MeV-100 GeV  energy band. The light curve shown in Figure~\ref{fermi3m} (left upper panel) indicates that 3C~120 was active starting from 2012. \cite{Abdo:2010vn} reported  also a detection between November 2008 and February 2009. Our new analysis, which was performed with up-dated background files and new instrument response functions, provides only a flux upper limit for the same time interval (second bin of the light curve). Although we can not claim a detection, the calculated TS value is however high (TS=9.2, corresponding to $\sim 3\sigma$) and  only  slightly below the usually adopted detection threshold (TS=10).

Considering  that PKS~0423$+$051 is at only 1.6$\degr$ from 3C120, that both sources have steep spectra and, that the LAT PSF is very broad at MeV energies, we decided to explore the 3-month light curve of the quasar to check for possible confusion effects. As shown in Figure~\ref{fermi3m}, PKS~0423$+$051 and 3C~120 flare in different periods and exhibit light curves with different pattern, suggesting a negligible (mutual) contamination. 

As further test, we also performed a variability study considering only photons  with energy $> 500$~MeV. Above this energy, the 68$\%$ containment angle (i.e., the radius of the circle containing 68$\%$ of the PSF) is indeed comparable or smaller than the separation of the two sources. The 500 MeV$-$100 GeV light curves (Figure~\ref{fermi3m}, right panel) are similar to those obtained taking also into account softer photons, supporting the previous conclusion on negligible confusion effects.

Finally, we reduced the integration time interval of each bin to 15 days to better constrain the time of the $\gamma$-ray detections, shown in Table~\ref{15d}. In this case the analysis has been extended until October 2014 to confirm the flare at the end of September 2014 reported in \cite{Tanaka:2014kx}. Our analysis yields a $\gamma-ray$ flux of $2.52\pm0.86\times10^{-8}$ photon cm$^{-2}$ s$^{-1}$ in MJD 56924-56939 for the September 2014 event (see Table~\ref{15d}).

\begin{table}[htpb]
    \begin{center}  
      \caption{$\gamma$-ray detections of 3C~120 in the energy range 0.5-100 GeV with a bin size of 15 days.\footnote{{\bf Notes.}We report the integration time in MJD and date, the flux with the corresponding error in 10$^{-8}$ photon cm$^{-2}$ s$^{-1}$ and the TS value associated to each $\gamma$-ray detection.}}
 \begin{tabular}{c c c c c} 
     \hline
     \hline
   MJD & Date &  Flux & Err & TS\\
   56264-56279 & 03/12/2012-18/12/2012 & 1.41 & 0.60 & 10.6\\
   56549-56564 & 14/09/2013-29/09/2013 & 1.65 & 0.54 & 21.6\\
   56564-56579 & 29/09/2013-14/10/2013 & 1.31 & 0.53 & 13.3\\
   56774-56789 & 27/04/2014-12/05/2014 & 1.15 & 0.52 & 10.4\\
   56819-56834 & 11/06/2014-26/06/2014 & 1.47 & 0.59 & 13.8\\
   56924-56939 & 24/09/2014-09/10/2014 & 2.52 & 0.86 & 18.4\\  
   \hline
     \end{tabular}
     \label{15d}
    % \footnotetext{}
 \end{center}
\end{table}

\section{The parsec scale jet at 15 and 43 GHz}
\subsection{VLBA data at 15 GHz}

\begin{figure*}[ht]
\centering
\includegraphics[width=1.34\textwidth]{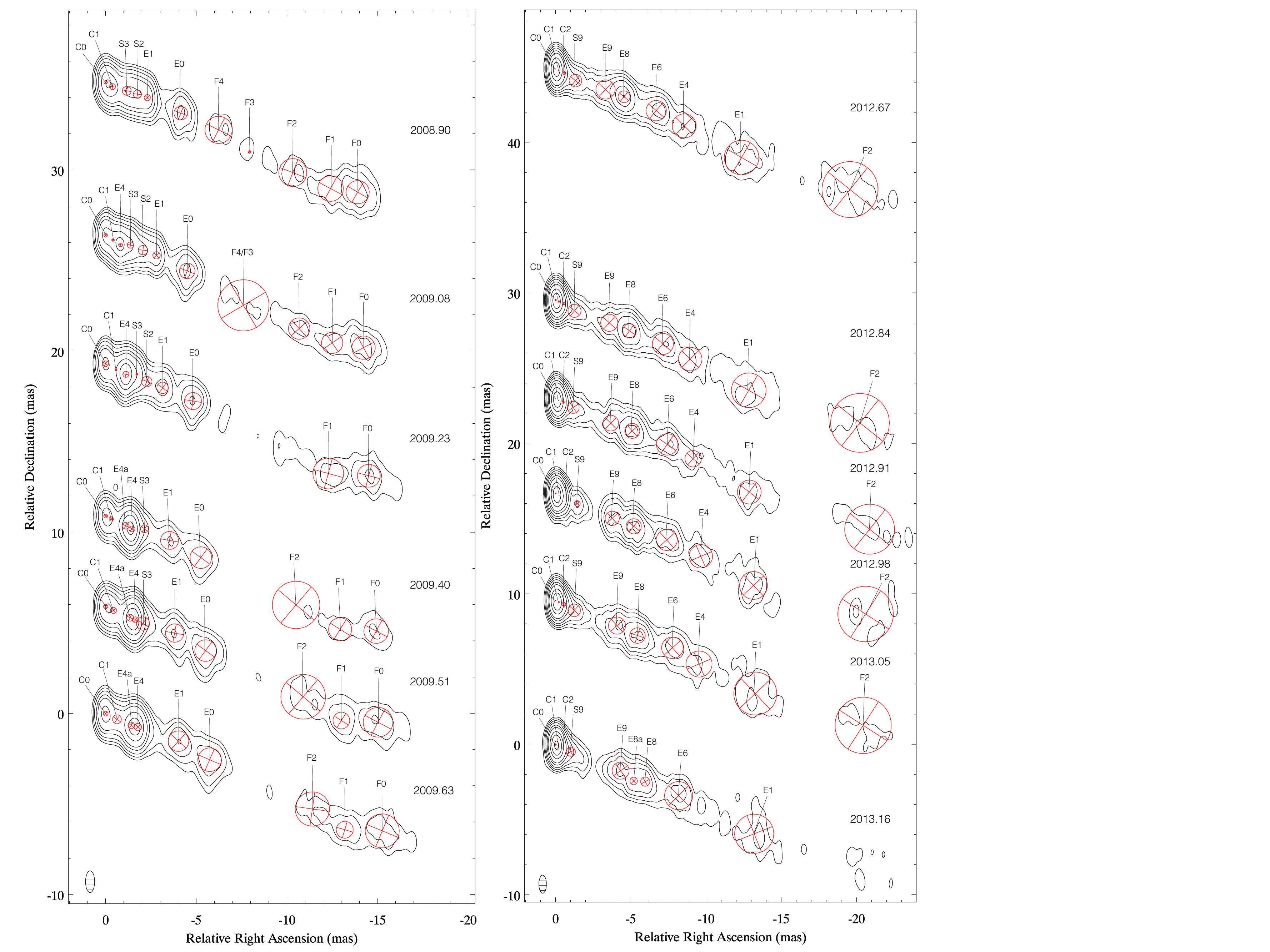}
\caption{Sequence of total intensity 15 GHz VLBA images from the MOJAVE monitoring program of 3C 120 with a common restoring beam of 1.2$\times$0.5 mas at 0$^{\circ}$. The separation among images is proportional to the time elapsed between observing epochs. Contours are traced at 0.0015, 0.004, 0.009, 0.02, 0.05, 0.1, 0.3, 0.6, 1.2 Jy. Red circles represent model-fit components.}
\label{fig:images_45_15G}
\end{figure*}

The study of the jet evolution at 15 GHz has been performed on the series of 46 VLBA images obtained by the MOJAVE monitoring program -- a subset of these images is displayed in Figure \ref{fig:images_45_15G}, where contours represent the total intensity with model-fit components (red circles) overlaid.   

We detected in total 25 components, apart from the core, although some of them characterize more probably the underlying flux density than knots that have been ejected from the core and move along the jet. The radio core is usually defined as the bright, compact feature at the upstream end of the jet, which may correspond to a recollimation shock (i.e., G\'omez et al. 1997) at millimeter wavelengths and to the optically thin-thick transition at centimeter wavelengths. 

\begin{figure}[htpb]
\centering
\includegraphics[width=0.49\textwidth]{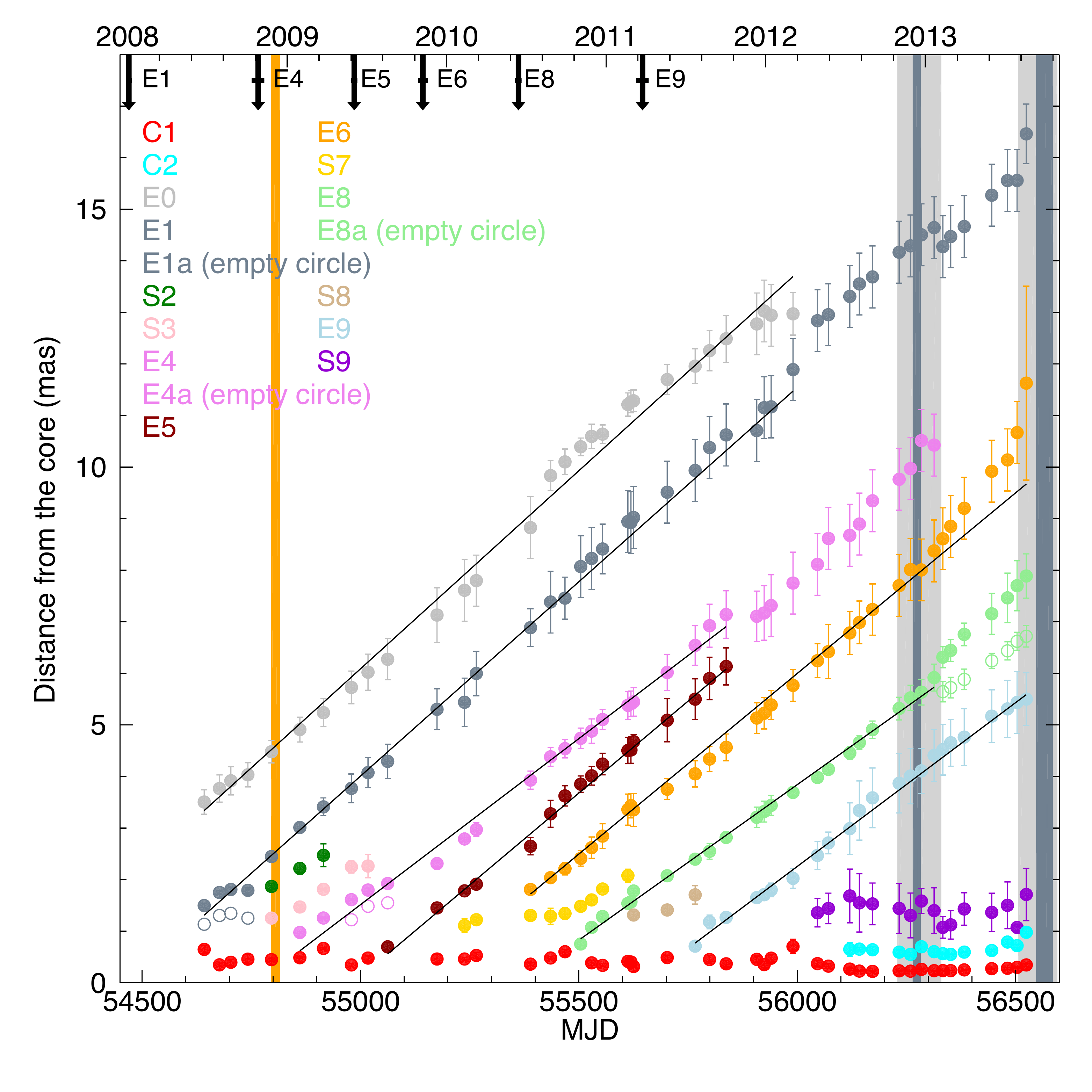}
\caption{Distance from the core vs. time for the 15 GHz model-fit components with linear fits overlaid. Downward black arrows mark the time of ejection of each component. Grey vertical lines indicate $\gamma$-ray detections in the 15 days-bin (dark grey) and 3 months-bin (light grey) light curves in the energy band 500 MeV-100 GeV. The orange vertical line marks the optical flare reported in \cite{Kollatschny:2014fk} (See text).}
\label{fig:Fit_15G}
\end{figure}

\begin{figure}[htpb]
\centering
\includegraphics[width=0.49\textwidth]{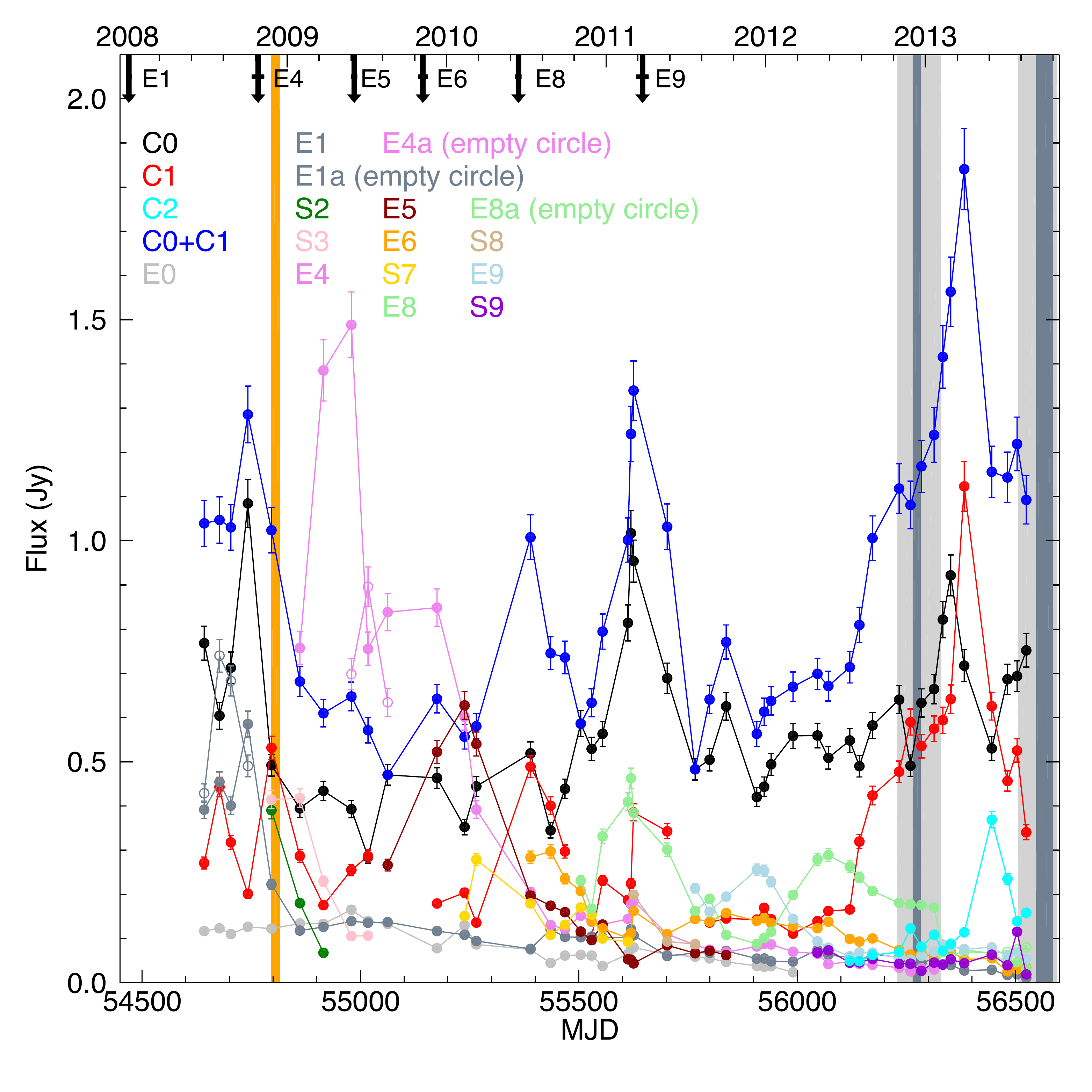}
\caption{Light curves of the 15 GHz model-fit components, including the added total flux density of the core and component C1 (C0+C1). Arrows and vertical lines are as indicated in Fig.~\ref{fig:Fit_15G}.}
\label{fig:Flux_15G}
\end{figure}

Components E-F and C are robustly identified moving and standing features, respectively, and the rest of model-fit components are required by the data, but cannot be cross-identified across the epochs and may represents e.g., emission from the underlying jet flow.  The core, identified with component C0, is considered stationary across epochs. Plots of the separation from the core and flux density evolution of the fitted components are shown in Figs.~\ref{fig:Fit_15G} and \ref{fig:Flux_15G}, respectively.

We fit the trajectories with respect to the core for all the superluminal knots that can be followed for a significant number of epochs, as plotted in Fig.~\ref{fig:Fit_15G}. In order to have a better determination of the time of ejection of each component, namely the time when a new knot crosses the radio core, we use linear fits for the component separation versus time. Note that in some cases we have found evidence for components merging, splitting, or a clear acceleration in their motion \citep[e.g.,][]{Homan:2009uq, Homan:2015uq}. In those cases we have considered only the initial epochs with a clear linear fit, as we are mainly interested in determining the time of ejection of each component. The time of ejection, angular and apparent velocities of the components are tabulated in Table~\ref{table_1}.

We find that superluminal components move with apparent velocities between 5 and 6 $c$, in agreement with previous findings \citep{Gomez:2001fk, Gomez:2008uq, Gomez:2011vn, Jorstad:2005fk}. This agrees also with recent MOJAVE results \citep{Lister:2013fk} where the 3C~120 kinematics, together with the radio galaxy 3C~111, departs from the others in the sample. These two radio~galaxies seem to be the only ones displaying clear superluminal motions and with apparent speeds that do not commonly change with distance from the core.

From June 2008 to the beginning of 2012 a stationary feature, C1, is found at a distance that shifts 
between $\sim$ 0.4 - 0.7 mas from the core. We identify this component with that reported by \cite{Leon-Tavares:2010uq} at 0.72$\pm$0.25 mas from the core, which is assumed by the authors to be related with optical flares when new components cross its position. Starting in 2012 the innermost 1 mas region of the jet changes, so that two stationary components can be found, C1 and C2. An extra component, labeled S9, is also required to fit some extended and weak flux density at $\sim$1.5 mas from the core.

As can be seen in Figs.~\ref{fig:images_45_15G} and \ref{fig:Flux_15G}, the presence of two stationary components (C1 and C2) within the innermost 1 mas is associated with an increase in flux density and a more extended structure of the core. Component C1 is observed to progressively increase its flux density between mid 2012 and the beginning of 2013, when both, the core and C1 are in a high flux density state. Three months later component C2 also shows an increase in flux density. These changes in the innermost structure of the jet are associated with the ejection of a new component, d11, revealed by the 43 GHz VLBA data (see section \ref{s:VLBA_43}), which extends our study of the jet until May 2014.

  Two other radio flares in the core have been also observed, one at the end of 2008 (2008.76$\pm$0.10 yr) and a second one at the beginning of 2011 (2011.15$\pm$0.05 yr). The flare in 2011 is associated with the ejection of a weak radio component from the core, E9, for which we estimate the time of ejection in 2011.23$\pm$0.04.   

\begin{table}[htpb]
  \begin{center}  
  \caption{Time of ejection, proper motions and apparent velocities of modelfit components at 15 GHz.}
  \begin{tabular}{lc c c c l} 
     \hline
     \hline
Name & $T_\mathrm{ej}$ & $\mu$ & $\beta_\mathrm{app}$ \\ 
  & (year) & (mas/yr)  \\ 
\hline
E0 & 2007.29 $\pm$ 0.06 & 2.81 $\pm$ 0.05 & 6.21 $\pm$ 0.11 \\
E1 & 2008.01 $\pm$ 0.02 & 2.76 $\pm$ 0.05 & 6.10 $\pm$ 0.11\\
E4 & 2008.82 $\pm$ 0.04 & 2.35 $\pm$ 0.05 & 5.19 $\pm$ 0.11\\
E5 & 2009.42 $\pm$ 0.02 & 2.60 $\pm$ 0.06 & 5.75 $\pm$ 0.13\\
E6 & 2009.85 $\pm$ 0.03 & 2.56 $\pm$ 0.05 & 5.66 $\pm$ 0.11\\
E8 & 2010.45 $\pm$ 0.02 & 2.20 $\pm$ 0.03 & 4.86 $\pm$ 0.07\\
E9 & 2011.23 $\pm$ 0.04 & 2.32 $\pm$ 0.09 & 5.12 $\pm$ 0.19\\
\hline
   \end{tabular} 
   \label{table_1}
  \end{center}
\end{table}

  The radio core flare at the end of 2008 is instead associated with the ejection of a bright superluminal knot, E4, in 2008.82$\pm$0.04 yr. The light curves of the different components shown in Fig.~\ref{fig:Flux_15G} reveal the unusually large flare experienced by E4. By mid 2009 it has doubled its flux density, reaching a peak of $\sim$ 1.5 Jy and becoming brighter than the core itself. During the high state of flux density of E4 another component, named E4a, appears very close to it, suggesting an increase in extension of component E4 during the flare, as can be seen in Figs.~\ref{fig:images_45_15G}, \ref{fig:Fit_15G}, and \ref{fig:Flux_15G}. In this case we consider more appropriate to take into account a flux-density-weighted distance between that of E4 and E4a to estimate the distance from the core of the new component E4 during its high state of flux density. The same method is used for components E1 and E1a, as we considered E1a an extension of E1 soon after this component is ejected from the core at the beginning of 2008. A similar splitting of components was also observed previously in 43 GHz VLBA images of 3C~120 \citep{Gomez:2001fk}
  
  The extended emission structure of components E1 and E4 (associated with components E1a and E4a) is consistent with \cite{Aloy:2003ys} relativistic hydrodynamic simulations, where the passage of a new perturbation from a series of recollimation shocks results in extended emitting regions due to light-travel time delays between the front and the back of the perturbation.

   Figure \ref{fig:Flux_15G} shows also the light-curve obtained from adding the flux densities of C0 and C1. The stationary component C1 is usually located at a distance from the core of the order or smaller than the observing beam, therefore in many epochs it is difficult to disentangle its flux density from that of the core. For instance, the combination of the C0+C1 flux density reveals a high state leading to the ejection of component E8 in mid 2010. It is also particularly remarkable the increase in flux density of the C0+C1 complex in mid 2013 leading to the ejection of a new component, d11, seen at 43 GHz (see below).

\subsection{VLBA data at 43 GHz}
\label{s:VLBA_43}

\begin{figure}[htbp]
\centering
\includegraphics[width=1.62\textwidth]{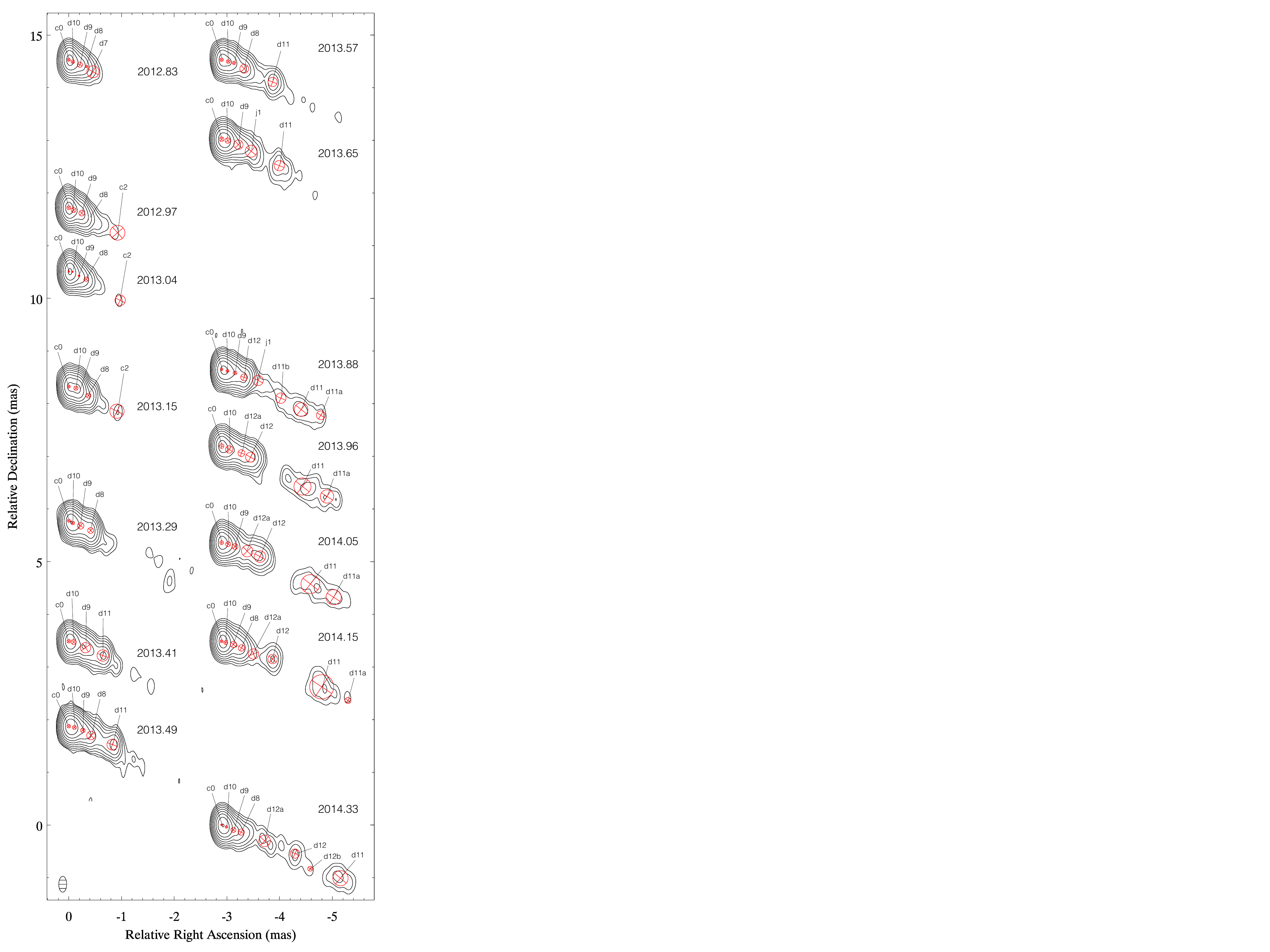}  
\caption{Sequence of total intensity 43 GHz VLBA images from the VLBA-BU-BLAZAR monitoring program of 3C~120 with a common restoring beam of 0.3$\times$0.15 mas at 0$^{\circ}$. Red circles represent modelfits components. Contours are traced at 0.16, 0.35, 0.77, 1 .7, 3.77, 8.33, 18.41, 40.71, 90\% of the peak intensity, 1.82 Jy/beam.}
\label{fig:images_43G}
\end{figure}

\begin{figure}[htbp]
\centering
\includegraphics[width=0.45\textwidth]{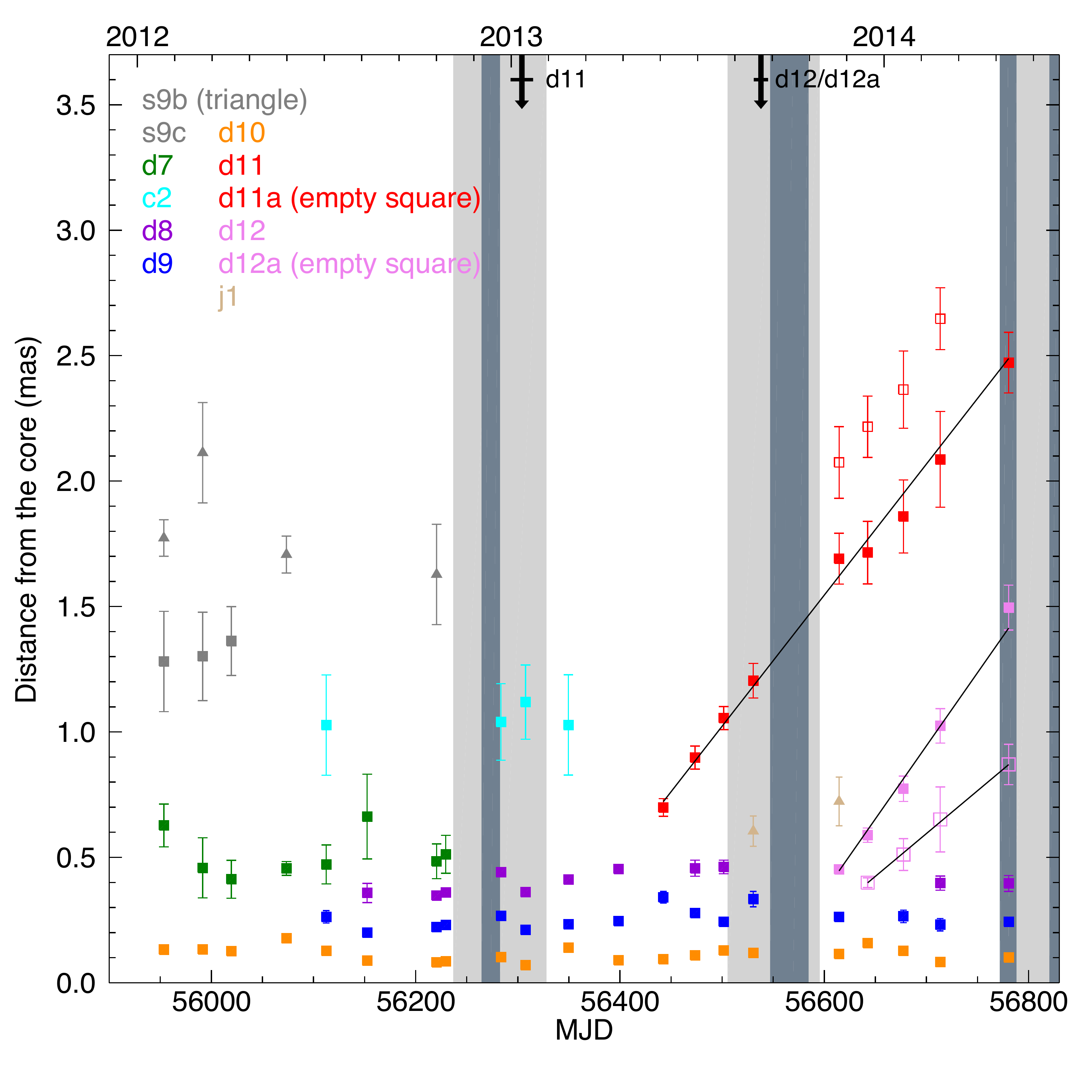}
\caption{Distance from the core vs. time for the 43 GHz model-fit components with linear fits overlaid. Downward black arrows mark the time of ejection of each component. Grey vertical lines indicate $\gamma$-ray detections in the 15 days-bin (dark grey) and 3 months-bin (light grey) light curves in the energy band 500 MeV-100 GeV.}
\label{fig:fit_43G}
\end{figure}

\begin{figure}[htbp]
\centering
\includegraphics[width=0.45\textwidth]{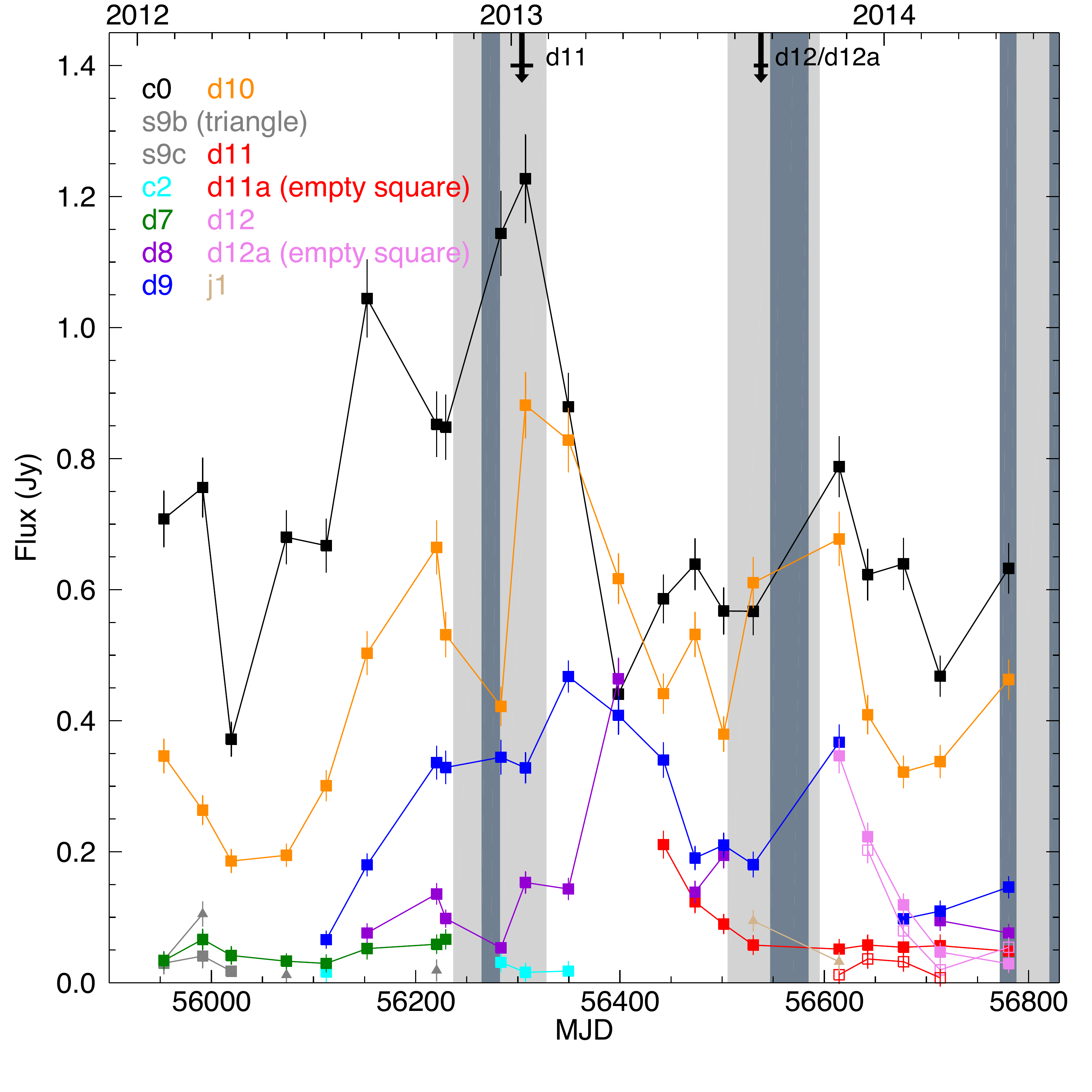}
\caption{Light curves of the 43 GHz model-fit components. Arrows and vertical lines are as indicated in Fig.~\ref{fig:fit_43G}.}
\label{fig:flux_43G}
\end{figure}

We modeled the 43 GHz jet in the same manner as for the 15 GHz data, finding a total of 16 components, although as in the case of the 15 GHz data, some of them are most probably related to the underlying continuum flow and it is difficult to follow them along the jet. In each epoch we identify the core with c0  and we consider it a stationary feature.

Figure \ref{fig:images_43G} shows a sequence of VLBA 43 GHz images in total intensity, where red circles represent model-fit components. The displayed images cover the observing period from October 2012 to May 2014, when 3C~120 was detected at $\gamma$-ray frequencies. 

  As observed in the 15 GHz data, starting in mid 2012 we find an increase in the flux density and extension of the core, requiring up to three stationary components, d10, d9, and d8, to model the innermost 0.5 mas structure (see Figs.~\ref{fig:fit_43G} and \ref{fig:flux_43G}). Because of the different angular resolution and opacity with respect to that obtained at 15 GHz, it is difficult to establish a one-to-one connection among the components at 43 and 15 GHz for the innermost jet region. We tentatively identify components s9b and s9c with substructures of component S9 seen at 15 GHz. Furthermore, during the enhanced activity of the source starting in mid 2012 it is not clear whether d10, d9, and d8 at 43 GHz -- and C1, C2 at 15 GHz -- correspond to actual physical structures in the jet, like recollimation shocks, or they trace the underlying emission of the jet.
  
  The radio core persists in a flaring state until January 2013. After this, the peak in flux density moves from the core along the jet, crossing progressively the three stationary features close to it (d10, d9 and d8). When the perturbation crosses the last stationary component, d8, we can clearly discern a new knot, d11, that emerges from the first 0.5 mas in 2013.4 yr (see Fig.~\ref{fig:images_43G}). From the progressive flaring of stationary features close to the core we can infer the extension of the crossing emitting region, obtaining $\sim$0.35 mas. This is significantly larger that the FWHM obtained from the model-fit of d11 (see Table~\ref{tab43G}), which suggests that d11 is in fact part of a more extended region, resembling the results obtained from relativistic hydrodynamical simulations \citep{Aloy:2003ys}. Fitting of the separation versus distance for d11 yields a proper motion of 1.91$\pm$0.09 mas/yr, which corresponds to an apparent velocity of 4.22$\pm$0.22{\it c}. The estimated time of ejection, that is, when component d11 crossed the radio core at 43 GHz, is 2013.03$\pm$0.03. 

A similar situation takes place also in the second half of 2013, when the core, together with components d10 and d9, starts to bright and a new component, d12, appears at $\sim$ 0.45 mas. We note also that another component, d12a, appears very close in time and position to component d12, although it displays a significantly slower proper motion. Component d12 moves at 2.1$\pm$0.2 mas/yr (4.7$\pm$0.3 $c$), while component d12a moves at 1.2$\pm$0.2 mas/yr (2.6$\pm$0.5 $c$). We note, however, that the estimated time of ejection for d12 and d12a, 2013.67$\pm$0.02 yr and 2013.64$\pm$0.07 yr, respectively, is the same within the uncertainties. Both components therefore originated simultaneously in the millimeter VLBI core, although they propagate at quite different velocities afterwards. 

Time delays stretch the shocked emission in the observer's frame, so that with the necessary angular resolution multiple sub-components associated with a single shock could be distinguished in the jet, but in this case they would have similar apparent velocities, in contrast with what it is observed for components d12 and d12a. Trailing components have a smaller velocity than the leading perturbation, but they are released on the wake of main perturbation \citep{Agudo:2001uq}, instead of being ejected from the core, as occurs for component d12a. On the other hand, relativistic hydrodynamic simulations show that a single perturbation in the jet inlet leads to the formation of a forward and reverse shock \citep{Gomez:1997kx,Aloy:2003ys,Mimica:2009fk}. Therefore, the fact that components d12 and d12a are ejected from the core at the same time but with different velocities suggests that they may correspond to the forward and reverse shock of a perturbation.

\section{Connection between $\gamma$-ray and radio emission}

  Our analysis of the {\it Fermi}-LAT data (see Figs.~\ref{fermi3m} and Tab.~\ref{15d}) shows a prolonged $\gamma$-ray activity in the radio~galaxy 3C~120 between December 2012 and September-October 2014, when the source reaches a flux of F$_{>500~\mathrm{MeV}} = 2.5\pm0.8\times10^{-8}$ photon cm$^{-2}$ s$^{-1}$, about an order of magnitude larger than in previous detections. Three clear periods of $\gamma$-ray activity are found at the end of 2012 (56264--56279 MJD), September-October 2013 (56549--56579 MJD), and May-October 2014 (56774--56939 MJD).

  The $\gamma$-ray activity in December 2012 is accompanied by an increase in the radio core flux density, as well as in the innermost stationary components, at both 15 and 43 GHz (see Figs.~\ref{fig:Flux_15G} and \ref{fig:flux_43G}), leading to the ejection of component d11 in 2013.13$\pm$0.03 (see also Fig.~\ref{fig:fit_43G}). From the estimated proper motion of d11 we infer that in the last epoch of the 15 GHz data this component is still in the innermost region of the jet as imaged at 15 GHz, crossing the stationary component C2.

  After this first $\gamma$-ray detection the source remains in a quiescent state until August 2013,   when another $\gamma$-ray event covering the period from August 2013 to October 2013 is detected. The 15-days binning analysis (see Tab.~\ref{15d}) constrains the main flaring activity to September-October 2013 (56549-56579 MJD), coincident with the flaring of the radio core at 43 GHz and the ejection of the d12-d12a pair of components (see Figs.~\ref{fig:images_43G} to \ref{fig:flux_43G}).
  
  There is also indication of the beginning of a new flaring activity in the 43 GHz radio core associated with the last $\gamma$-ray flaring activity starting in May 2014, but further VLBI images are required to confirm the 
  ejection of a new component in this event.
  
  We therefore can conclude that there is a clear association between the $\gamma$-ray and radio emission in 3C~120, such that every period of $\gamma$-ray activity is accompanied by flaring of the VLBI radio core and subsequent ejection of a new superluminal component in the jet. However, not all ejections of superluminal components are related to enhanced $\gamma$-ray emission, detectable by {\it Fermi}-LAT, as occurred for components E4, E5, E6, E8, and E9. The case of component E4 is particularly interesting. \cite{Abdo:2010vn} report a $\gamma$-ray detection in December 2008, coincident with an increase of the 15 GHz radio core flux, leading to the ejection of E4 in 2008.82$\pm$0.04. This is also coincident with an optical flare in December 2008, as reported by \cite{Kollatschny:2014fk}. We also note that E4 reached a peak flux of $\sim$ 1.5 Jy, significantly larger than any other component seen in our analysis. Despite this intense activity in the optical and radio core, our analysis of the {\it Fermi} data during this period does not provide a clear detection in $\gamma$-rays, although the calculated TS (corresponding to $\sim 3\sigma$) 
  is only slightly smaller than the usually adopted detection threshold.

\section{Discussion}

\subsection{Motion of components along the jet}

The debate on the origin and location of the $\gamma$-ray emission in blazars has gained added interest since the launch of the {\emph Fermi} satellite. Much of the current discussion lies in whether $\gamma$-rays are produced upstream of the mm-VLBI core, as suggested by some $\gamma$-ray and radio correlations~\citep[e.g.][]{Rani:2013uq, Rani:2014fk} and the observed $\gamma$-ray spectral break at few GeVs~\citep{Abdo:2009vn, Finke:2010ys, Poutanen:2010fk, Tanaka:2011zr, Rani:2013kx} or downstream, as suggested by coincidence of $\gamma$-ray flares with either the appearance of new superluminal components~\citep[e.g.][]{Jorstad:2010uq, Jorstad:2013wd} or the passage of moving components through a stationary jet feature \citep{Schinzel:2012qf, Marscher:2013ve}.

 As observed in several blazars \citep{Jorstad:2010uq, Jorstad:2013wd, Ramakrishnan:2014kx}, the interaction between traveling features and the stationary radio core appears to be a necessary condition for the production of $\gamma$-ray photons in 3C~120, but it is clearly not enough. Therefore, to understand the $\gamma$-ray emission in 3C~120, and more generally in AGN, it is necessary to address the question of what physical changes in the jet can produce $\gamma$-ray emission.

  We note that the beginning of the $\gamma$-ray activity in 3C~120 occurs after a sustained period of low activity in the jet. As indicated in Table~\ref{table_1}, during the time period analyzed new components are seen in the jet of 3C~120 roughly every 8 months; however, no new components are detected in the jet between the ejection of E9 in 2011.23$\pm$0.04 and component d11 in 2013.03$\pm$0.03, implying a lack of activity in the jet for almost 2 years. Note that after the ejection of d11 the source resumes its activity with the ejection of component d12, again roughly 8 months later. 
  
 % \clearpage
\begin{figure*}[htbp]
\centering
\includegraphics[width=0.49\textwidth]{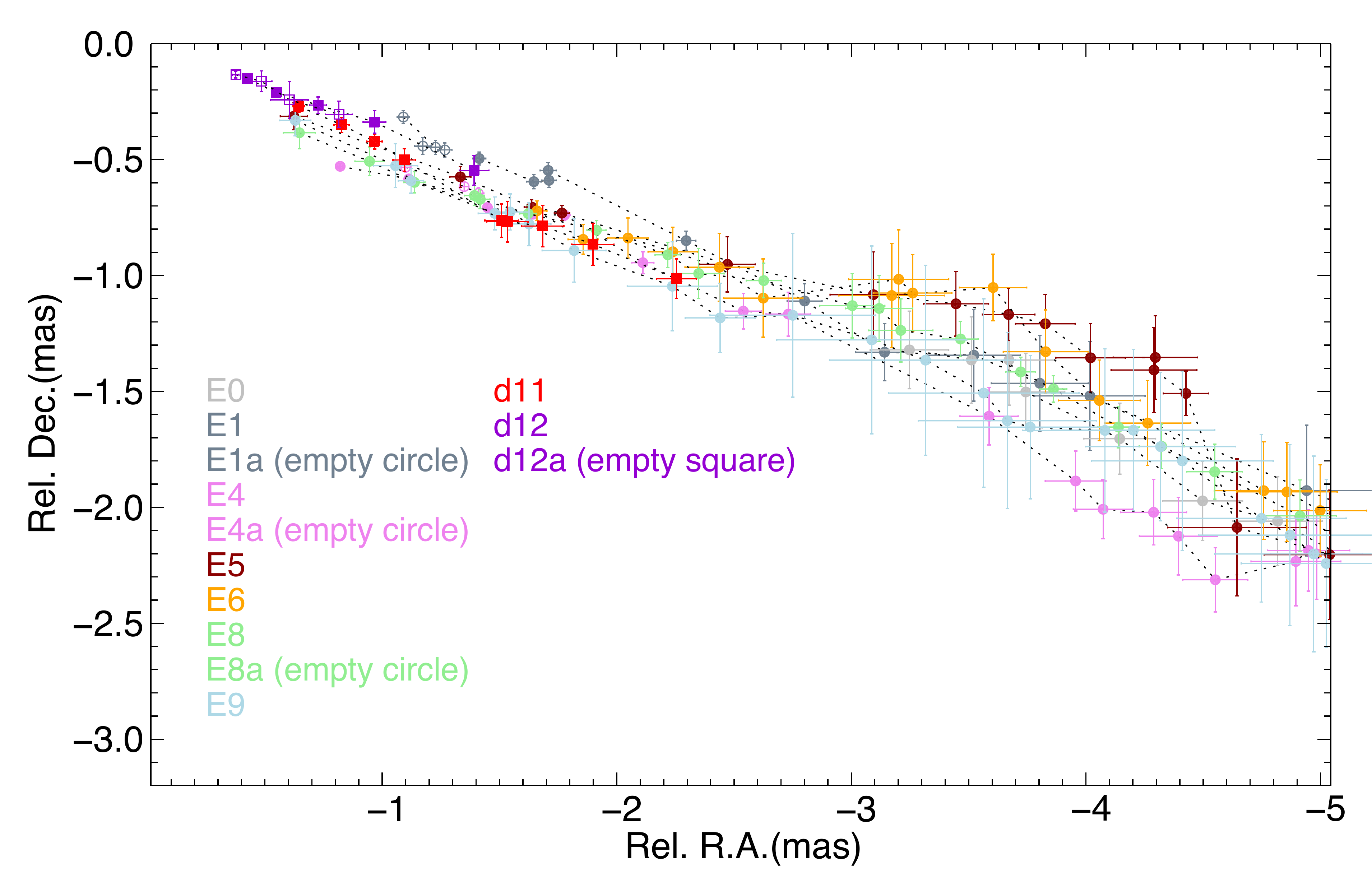}
\includegraphics[width=0.49\textwidth]{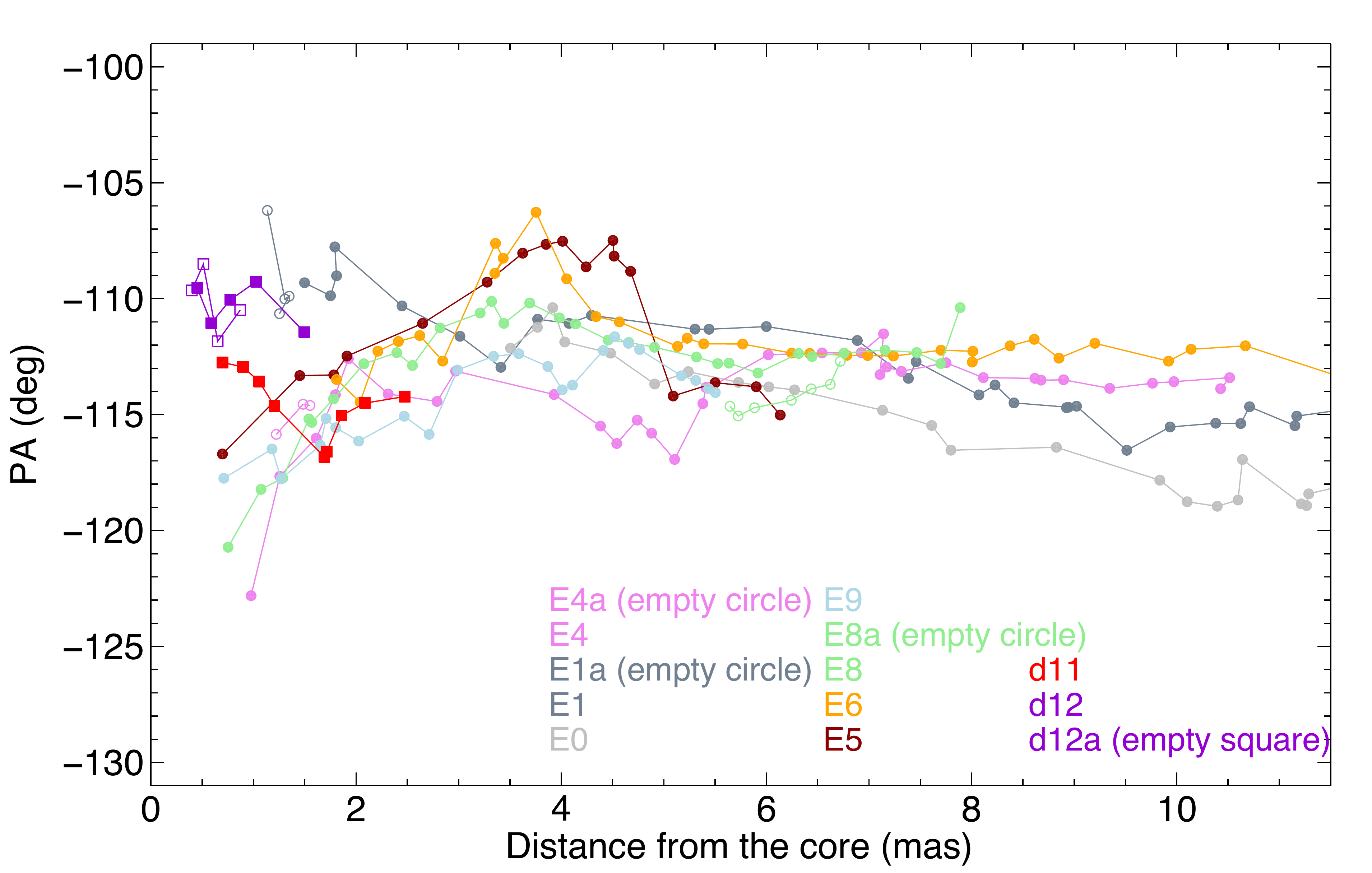}
\caption{Positions ({\emph left}) and position angles as a function of distance from the core ({\emph right}) for the model-fit components at 15 (circles) and 43 GHz (squares). For clarity errors bars are shown only in the plot of the components' position.}
\label{fig:PA_Dist}
\end{figure*}

  Analysis of the components' proper motions (see Table \ref{table_1}) reveals a clear pattern of decreasing apparent velocities, from the 6.21$\pm$0.11$c$ of E0 to 5.12$\pm$0.19$c$ of E9. Furthermore, when the core resumes the ejection of components after a 2-year inactivity, the apparent velocities measured for d11 and d12 have further decreased to 4.22$\pm$0.22$c$ and 4.70$\pm$0.31$c$, respectively. Component d12a shows an even smaller velocity of 2.75$\pm$0.45$c$, but we believe this component differs for the other components seen previously in 3C~120 in which it is probably associated with a reserve shock.

  Note that model-fits at 43 GHz and 15 GHz trace the motion of components continuously throughout the jet. We also find no evidence for acceleration (only component E6 shows a clear acceleration at $\sim$8 mas from the core), hence we consider that the proper motions measured at 43 GHz provide a good estimation for the expected values should these components be detected later on at 15 GHz, reassuring our finding for a progressive decrease in the apparent velocity of the components from $\sim$6.2$c$ to 4.2$c$ in a time span of approximately 6.4 years, from ~2007.3 to 2013.7.  This progressive change in the apparent velocity of components could be due to a change in the velocity and/or orientation of the components. Considering that the 6.4 yr time span measured roughly agrees with the 12.3$\pm$0.3 yr full period determined by the precessing jet model of \cite{Caproni:2004fk}, we favor the case in which it is produced by a change in the orientation of the components with respect to the observer.

  Figure \ref{fig:PA_Dist} shows the motion of components along the jet, as well as their position angle (PA) with distance from the core. We observe that components E1-E1a (and presumably E0) are ejected with a PA of around $-110^{\circ}$ and they travel towards the south-west direction while the other components at 15 GHz are ejected with a PA between $-117^{\circ}$ and $-123^{\circ}$ and move initially in a less southern direction. Components E5 and E6 present a significant change in their velocity vectors at a distance of $\sim$3 mas from the core, as observed previously for other components and interpreted as the interaction with the external medium or a cloud (G\'omez et al. 2000, 2008). Components d11 and d12-d12a are ejected with a PA of around $-110^{\circ}$, similar to that observed previously for E1-E1a. Note that in a precession model components E1 and d12 would have similar projected velocity vectors while differing in their orientation with respect to the observer. Despite the initially different position angles, components move following parallel paths after the initial $\sim$ 4-5 mas. Rather than a precession of the whole jet, we therefore favor the model in which the jet consists of a broad funnel through which components -- not filling the whole jet width -- are ejected and travel at differ position angles, as supported also by previous observations of 3C~120 \citep{Gomez:2011vn} and other sources in the MOJAVE sample \citep{Lister:2013fk}.

  Considering that the slower apparent velocities of d11 and d12 are related to the first $\gamma$-ray detections, we conclude that these components are most likely ejected with a smaller viewing angle. This should increase the Doppler factor, leading to the enhanced $\gamma$-ray emission measured since the end of 2012, and a significant increase in the total flux of the core and innermost stationary features (see Fig.~\ref{fig:flux_43G}). However, we cannot completely rule out the possibility that the smaller apparent velocities are just due to greater viewing angles, or slowing of the components' velocity.
  
  It is possible to estimate the required minimum Lorentz factor, $\Gamma_\mathrm{min}=(1+ \beta_{app}^2)^{1/2}=6.3$, using the observed maximum apparent velocity of 6.2$c$. To minimize the required reorientation of the components, we can assume that the maximum apparent velocity is obtained for the angle that actually maximizes the apparent velocity, given by $\theta=\arccos(\beta)$, where $\beta$ is the component's velocity in units of the speed of light. Using our previous estimation of $\Gamma_\mathrm{min}=6.3$, we obtain a viewing angle of $\theta \sim 9.2^{\circ}$ for the maximum apparent velocity of 6.2$c$ measured for component E0, in agreement with \cite{Hovatta:2009fk}. Given that the observed decrease in the apparent velocity is probably produced by a decrease in the viewing angle, the smallest observed apparent velocity of 4.2$c$ for d11 requires $\theta\sim3.6^{\circ}$. These values agree with the precession model of Caproni \& Abraham (2004), for which the authors estimate $\Gamma=6.8\pm0.5$ and a variation of the viewing angle between (6.3$\pm$0.8$^{\circ}$) and (3.3$\pm$0.8$^{\circ}$). We should note however that our measured change in the apparent velocities is shifted from the periodicity phase predicted by \cite{Caproni:2004fk} model.
    
  This change in the orientation from $\theta \sim 9.2^{\circ}$ to $\theta\sim3.6^{\circ}$ would lead to an increase in the Doppler factor from $\delta \sim 6.2$ to $\delta \sim 10.9$, enhancing the $\gamma$-ray emission above the flux detectable by {\emph Fermi}.

  In summary, all these evidence suggest that the observed $\gamma$-ray emission in 3C~120 depends strongly on the orientation of component's motion with respect to the observer, so that only when they are best oriented {\it and} a new superluminal component pass through the radio core a clear $\gamma$-ray detection is obtained.
  
   Superluminal components are associated with shocks moving along the jet, hence the inferred values for the velocity -- provided there is an estimation of the jet orientation -- correspond to the pattern velocity of the shock, not the actual flow velocity. The detection of the forward and reserve shocks of the perturbation associated with d12 and d12a, respectively, allows to obtain a direct calculation of the jet bulk flow velocity. For the estimated viewing angle of $\sim3.6^{\circ}$ during the ejection of components d12 and d12a we obtain the corresponding pattern velocities of $\Gamma_\mathrm{d12}$=6.7 and $\Gamma_\mathrm{d12a}$=5. Hence, we can conclude that the jet bulk flow velocity is restricted to be $5\leq\Gamma_\mathrm{j}\leq6.7$.

\subsection{Gamma-ray location and emission mechanism}

Our first $\gamma$-ray detection at the end of 2012 is associated with the passage of component d11 through the mm-VLBI core, whose time of ejection is coincident with the 3 months-bin $\gamma$-ray detection, and is $\sim$ 34 days (0.09 yr) after the 15 days-bin $\gamma$-ray detection. On the other hand, \cite{Marscher:2002kx} and \cite{Chatterjee:2009ve} have measured a mean delay of 0.18 yr ($\sim$ 66 days) between X-ray dips and the ejection of new superluminal components. Considering that most of the X-ray emission in 3C~120 originates from the disk-corona system, close to the central black hole (BH), we can consider this time delay as the distance in time between the BH and the mm-VLBI core. A mean apparent motion of $\sim$ 2 mas/yr corresponds to a rate of increase in projected separation from the core of $\sim$0.24 pc/yr -- or slightly smaller if we allow for some initial acceleration of the components. 

The measured delay of 34 days between the 15 days-bin $\gamma$-ray detection and the passage of d11 from the mm-VLBI core locates the $\gamma$-ray emission upstream of the mm-VLBI core, at a distance about half of that between the BH and the core. Relativistic 3D hydrodynamic simulations of \cite{Aloy:2003ys} show that time delays stretch significantly the observed size of components, so that our VLBI observations could detect only some portions of the component, depending on the strength of the forward/reverse shock and Doppler factor (i.e., viewing angle). Considering also that we have estimated for component d11 a size of $\sim$0.35 mas (0.2 pc) once it is clearly detached from the core, it is possible that the 15 days-bin $\gamma$-ray detection could correspond to the passing of the forward section of the d11 perturbation through the mm-VLBI core, while later on our VLBI images identify only the back section, as it is precisely seen in the numerical simulations of \cite{Aloy:2003ys}. In this case the flaring in $\gamma$-rays would more closely mark the crossing of the d11 perturbation through the mm-VLBI core.

The second $\gamma$-ray event is instead associated with the ejection of components d12 and d12a. In this case the time of ejection of these two components is also coincident with the 3 months bin $\gamma$-ray detection, but the finer time sampling of the 15-days $\gamma$-ray light curve constrains the $\gamma$-ray production to $\sim$ 33 days after. Hence we conclude that in the second $\gamma$-ray event the high energy emission is produced downstream of the mm-VLBI core, at a projected distance of $\sim$ 0.13 pc from its position -- smaller if we consider that the shock responsible for the $\gamma$-ray emission is the slower moving d12a.

Reverberation mapping studies of the broad-line-region (BLR) in 3C~120 suggest an inclined disk model with an extension from 12$\pm$7 light days ($\sim$0.01 pc) to 28$\pm$9 light days ($\sim$0.025 pc), from the BH \citep{Grier:2013uq, Kollatschny:2014fk}. These studies also found evidence of radial stratification in the BLR as well as infall and rotation related to the BH gravity. Therefore the BLR extends to about half of the estimated distance between the BH and the mm-VLBI core, severely limiting the external photon field from the BLR at the location of the mm-VLBI core. Hence we favor synchrotron self Compton as the mechanism for the production of $\gamma$-ray photons in the case of the second $\gamma$-ray event in late 2013, although we cannot discard the contribution from another external photon field, such the sheath or the external ionized cloud. 

  The External Compton process with photons coming from the BLR could be instead a possibility for the first $\gamma$-ray event, as our findings point to a $\gamma$-ray dissipation zone between the BH and the radio core, on the limit of the extension of the BLR.
  
  The different orientation of the components within the broader jet funnel supports also a model in which they interact with the sheath of the jet \citep[see also][]{Gomez:2000ys,Gomez:2008uq}. In this case the observed $\gamma$-ray activity can be produced by Compton scattering of photons from the jet sheath, as proposed by \cite{Marscher:2010uq} to explain the $\gamma$-ray activity seen in PKS~1510-089. 
  
\cite{Grandi:2012fk} obtain similar results for the FRII radio~galaxy 3C~111, associating the $\gamma$-ray activity with the ejection of a new radio component from the core, and confining the $\gamma$-ray dissipation region within 0.1 pc at a distance of almost 0.3 pc from the black hole. These two radio~galaxies have other similarities as they are both BLRGs and they show a connection between the radio jet and the corona-disk system \citep{Marscher:2002kx, Chatterjee:2011qf}. In addition, the apparent motions of superluminal components detected in their jets detach from those of the other radio~galaxies in the MOJAVE sample \citep{Lister:2013fk}.

\acknowledgements This research has been supported by the Spanish Ministry of Science and Innovation grants AYA2010-14844, and AYA2013-40825-P, and by the Regional Government of Andaluc\'{\i}a (Spain) grant P09-FQM-4784. The VLBA is an instrument of the National Radio Astronomy Observatory, a facility of the National Science Foundation of the USA operated under cooperative agreement by Associated Universities, Inc. (USA). This work has made use of data from the MOJAVE database that is maintained by the MOJAVE team \citep{Lister:2009vn}. This research was partly supported by the Russian Foundation for Basic Research grant 13-02-12103 and by the Academy of Finland project 274477.

%% Bibliography

\bibliography{3c120_Casadio}%{}

\begin{center}
 \framebox[\textwidth]{{\color{blue} Table \ref{tab15G}} and {\color{blue} \ref{tab43G}} are available in the electronic edition of the journal}\par 
\end{center}

% Electronic tables with the physical parameters of each component
%___________________________________________________________________

%____________________________________________________________________________
%____________________________ VLBA-15 GHZ____________________________________
%____________________________________________________________________________

\clearpage
\newpage

\begin{longtable*}{c c c c c c c c c c}
\caption[]{- VLBA 15 GHz Model-fit Components' Parameters}
\endfirsthead
%\hline\\ 
%\multicolumn{7}{|c|}{} \\ [0.07cm]
\hline\\ [0.07cm] 
Epoch & Epoch & Name & Flux  & Distance from & Pos. Angle & FWHM  \\ [0.07cm]
(year) & (MJD) & & (mJy) & C0 (mas) & ($^{\circ}$) &  (mas)\\ [0.07cm] 
\hline\\ [0.07cm]
2008.48 & 54642.50 & C0 & 768 $\pm$ 38 &&& 0.31 $\pm$ 0.20 \\ [0.07cm] 
&& C1  & 271 $\pm$ 14 & 0.65 $\pm$ 0.08 & $-$115.2 $\pm$ 6.7 & 0.49 $\pm$ 0.16 \\ [0.07cm] 
&& E0 & 117 $\pm$ 6 & 3.51 $\pm$ 0.24 & $-$112.1 $\pm$ 3.6 & 0.79 $\pm$ 0.47 \\ [0.07cm] 
&& E1 & 392 $\pm$ 20 & 1.50 $\pm$ 0.04 & $-$109.3 $\pm$ 1.4 & 0.32 $\pm$ 0.08 \\ [0.07cm] 
&& E1a & 429 $\pm$ 21 & 1.14 $\pm$ 0.04 & $-$106.2 $\pm$ 1.7 & 0.33 $\pm$ 0.08 \\ [0.07cm] 
\hline\\ [0.07cm]  
2008.58 & 54677.50 & C0 & 604 $\pm$ 30 &&& 0.22 $\pm$ 0.17 \\ [0.07cm] 
&& C1 & 443 $\pm$ 22 & 0.35 $\pm$ 0.03 & $-$115.8 $\pm$ 5.3 & 0.31 $\pm$ 0.07 \\ [0.07cm] 
&& E0 & 123 $\pm$ 6 & 3.77 $\pm$ 0.26 & $-$111.2 $\pm$ 3.7 & 0.88 $\pm$ 0.53 \\ [0.07cm] 
&& E1 & 455 $\pm$ 23 & 1.75 $\pm$ 0.04 & $-$109.9 $\pm$ 1.3 & 0.38 $\pm$ 0.09 \\ [0.07cm] 
&& E1a & 740 $\pm$ 37 & 1.31 $\pm$ 0.04 & $-$110.0 $\pm$ 1.7 & 0.48 $\pm$ 0.09 \\ [0.07cm] 
\hline\\ [0.07cm] 
2008.65 & 54703.50 & C0 & 713 $\pm$ 36 &&& 0.22 $\pm$ 0.18 \\ [0.07cm] 
&& C1 & 318 $\pm$ 16 & 0.40 $\pm$ 0.06 & $-$113.8 $\pm$ 7.9 & 0.41 $\pm$ 0.12 \\ [0.07cm] 
&& E0 & 110 $\pm$ 6 & 3.92 $\pm$ 0.27 & $-$110.4 $\pm$ 3.7 & 0.87 $\pm$ 0.55 \\ [0.07cm] 
&& E1 & 401 $\pm$ 20 & 1.81 $\pm$ 0.04 & $-$109.0 $\pm$ 1.2 & 0.35 $\pm$ 0.09 \\ [0.07cm] 
&& E1a & 683 $\pm$ 34 & 1.35 $\pm$ 0.04 & $-$109.9 $\pm$ 1.6 & 0.46 $\pm$ 0.09 \\ [0.07cm] 
\hline\\ [0.07cm] 
2008.76 & 54742.50 & C0 & 1084 $\pm$ 54 &&& 0.31 $\pm$ 0.12 \\ [0.07cm] 
&& C1 & 201 $\pm$ 10 & 0.46 $\pm$ 0.09 & $-$116.3 $\pm$ 10.1 & 0.46 $\pm$ 0.18 \\ [0.07cm] 
&& E0 & 127 $\pm$ 6 & 4.03 $\pm$ 0.24 & $-$111.9 $\pm$ 3.1 & 0.82 $\pm$ 0.47 \\ [0.07cm] 
&& E1 & 586 $\pm$ 29 & 1.79 $\pm$ 0.05 & $-$107.8 $\pm$ 1.4 & 0.47 $\pm$ 0.10 \\ [0.07cm] 
&& E1a & 491 $\pm$ 25 & 1.25 $\pm$ 0.05 & $-$110.7 $\pm$ 2.2 & 0.45 $\pm$ 0.10 \\ [0.07cm] 
\hline\\  [0.07cm] 
2008.90 & 54796.50 & C0 & 492 $\pm$ 25 &&& 0.18 $\pm$ 0.09 \\ [0.07cm] 
&& C1 & 532 $\pm$ 27 & 0.45 $\pm$ 0.03 & $-$122.6 $\pm$ 4.2 & 0.33 $\pm$ 0.07 \\ [0.07cm] 
&& E0 & 122 $\pm$ 6 & 4.48 $\pm$ 0.22 & $-$112.4 $\pm$ 2.6 & 0.75 $\pm$ 0.43 \\ [0.07cm] 
&& E1 & 223 $\pm$ 11 & 2.45 $\pm$ 0.06 & $-$110.3 $\pm$ 1.2 & 0.34 $\pm$ 0.12 \\ [0.07cm] 
&& S2 & 390 $\pm$ 20 & 1.87 $\pm$ 0.06 & $-$110.6 $\pm$ 2.6 & 0.45 $\pm$ 0.12 \\ [0.07cm] 
&& S3 & 415 $\pm$ 21 & 1.25 $\pm$ 0.06 & $-$112.1 $\pm$ 2.7 & 0.49 $\pm$ 0.13 \\ [0.07cm] 
\hline\\ [0.07cm] 
2009.08 & 54861.50 & C0 & 395 $\pm$ 20 &&& 0.19 $\pm$ 0.10 \\ [0.07cm] 
&& C1 & 287 $\pm$ 14 & 0.49 $\pm$ 0.02 & $-$123.0 $\pm$ 3.0 & 0.14 $\pm$ 0.05 \\ [0.07cm] 
&& E0 & 134 $\pm$ 7 & 4.91 $\pm$ 0.24 & $-$113.7 $\pm$ 2.6 & 0.86 $\pm$ 0.48 \\ [0.07cm] 
&& E1 & 118 $\pm$ 6 & 3.01 $\pm$ 0.11 & $-$111.6 $\pm$ 1.8 & 0.40 $\pm$ 0.21 \\ [0.07cm] 
&& E4 & 757 $\pm$ 38 & 0.98 $\pm$ 0.02 & $-$122.8 $\pm$ 0.9 & 0.22 $\pm$ 0.03 \\ [0.07cm] 
&& S2 & 180 $\pm$ 9 & 2.22 $\pm$ 0.11 & $-$111.9 $\pm$ 2.6 & 0.52 $\pm$ 0.22 \\ [0.07cm] 
&& S3 & 418 $\pm$ 21 & 1.47 $\pm$ 0.04 & $-$111.7 $\pm$ 3.0 & 0.35 $\pm$ 0.08 \\ [0.07cm] 
\hline\\  [0.07cm] 
2009.23 & 54915.50 & C0 & 434 $\pm$ 22 &&& 0.31 $\pm$ 0.15 \\ [0.07cm] 
&& C1 & 176 $\pm$ 9 & 0.67 $\pm$ 0.02 & $-$120.6 $\pm$ 1.5 & 0.11 $\pm$ 0.05 \\ [0.07cm] 
&& E0 & 133 $\pm$ 7 & 5.24 $\pm$ 0.27 & $-$113.2 $\pm$ 2.7 & 0.94 $\pm$ 0.54 \\ [0.07cm] 
&& E1 & 126 $\pm$ 6 & 3.41 $\pm$ 0.17 & $-$112.9 $\pm$ 2.7 & 0.63 $\pm$ 0.35 \\ [0.07cm] 
&& E4 & 1385 $\pm$ 69 & 1.26 $\pm$ 0.02 & $-$117.7 $\pm$ 0.8 & 0.33 $\pm$ 0.05 \\ [0.07cm] 
&& S2 & 68 $\pm$ 3 & 2.47 $\pm$ 0.22 & $-$113.4 $\pm$ 4.8 & 0.57 $\pm$ 0.44 \\ [0.07cm] 
&& S3 & 230 $\pm$ 11 & 1.81 $\pm$ 0.01 & $-$108.7 $\pm$ 1.5 & 0.11 $\pm$ 0.05 \\ [0.07cm] 
\hline\\ [0.07cm] 
2009.40 & 54979.50 & C0 & 393 $\pm$ 20 &&& 0.21 $\pm$ 0.09 \\ [0.07cm] 
&& C1 & 255 $\pm$ 13 & 0.35 $\pm$ 0.03 & $-$117.1 $\pm$ 5.5 & 0.23 $\pm$ 0.07 \\ [0.07cm] 
&& E0 & 165 $\pm$ 8 & 5.73 $\pm$ 0.32 & $-$113.6 $\pm$ 3.0 & 1.19 $\pm$ 0.63 \\ [0.07cm] 
&& E1 & 139 $\pm$ 7 & 3.77 $\pm$ 0.28 & $-$110.9 $\pm$ 3.9 & 0.98 $\pm$ 0.56 \\ [0.07cm] 
&& E4 & 1489 $\pm$ 74 & 1.61 $\pm$ 0.02 & $-$116.0 $\pm$ 0.6 & 0.32 $\pm$ 0.04 \\ [0.07cm] 
&& E4a & 698 $\pm$ 35 & 1.22 $\pm$ 0.03 & $-$115.9 $\pm$ 1.4 & 0.37 $\pm$ 0.06 \\ [0.07cm] 
&& S3 & 106 $\pm$ 5 & 2.25 $\pm$ 0.12 & $-$108.2 $\pm$ 2.8 & 0.43 $\pm$ 0.24 \\ [0.07cm] 
\hline\\ [0.07cm] 
2009.51 & 55017.50 & C0 & 285 $\pm$ 14 &&& 0.24 $\pm$ 0.09 \\ [0.07cm] 
&& C1 & 287 $\pm$ 14 & 0.48 $\pm$ 0.05 & $-$116.4 $\pm$ 5.7 & 0.34 $\pm$ 0.10 \\ [0.07cm] 
&& E0 & 140 $\pm$ 7 & 6.02 $\pm$ 0.35 & $-$113.8 $\pm$ 3.1 & 1.19 $\pm$ 0.70 \\ [0.07cm] 
&& E1 & 136 $\pm$ 7 & 4.08 $\pm$ 0.29 & $-$111.1 $\pm$ 3.8 & 1.01 $\pm$ 0.58 \\ [0.07cm] 
&& E4 & 756 $\pm$ 38 & 1.80 $\pm$ 0.03 & $-$114.1 $\pm$ 0.8 & 0.33 $\pm$ 0.05 \\ [0.07cm] 
&& E4a & 896 $\pm$ 45 & 1.48 $\pm$ 0.03 & $-$114.6 $\pm$ 1.1 & 0.38 $\pm$ 0.06 \\ [0.07cm] 
&& S3 & 107 $\pm$ 5 & 2.26 $\pm$ 0.23 & $-$114.5 $\pm$ 5.4 & 0.73 $\pm$ 0.45 \\ [0.07cm] 
\hline\\  [0.07cm]
2009.63 & 55062.50 & C0 & 470 $\pm$ 24 &&& 0.26 $\pm$ 0.08 \\ [0.07cm]
&& E0 & 133 $\pm$ 7 & 6.27 $\pm$ 0.40 & $-$113.9 $\pm$ 3.4 & 1.30 $\pm$ 0.79 \\ [0.07cm]
&& E1 & 137 $\pm$ 7 & 4.29 $\pm$ 0.33 & $-$110.7 $\pm$ 4.5 & 1.13 $\pm$ 0.67 \\ [0.07cm]
&& E4 & 839 $\pm$ 42 & 1.93 $\pm$ 0.03 & $-$112.6 $\pm$ 0.9 & 0.41 $\pm$ 0.07 \\ [0.07cm]
&& E4a & 635 $\pm$ 32 & 1.55 $\pm$ 0.03 & $-$114.6 $\pm$ 1.2 & 0.36 $\pm$ 0.07 \\ [0.07cm]
&& E5 & 266 $\pm$ 13 & 0.70 $\pm$ 0.08 & $-$116.7 $\pm$ 3.1 & 0.49 $\pm$ 0.16 \\ [0.07cm]
\hline\\  [0.07cm]
2009.94 & 55175.50 & C0 & 463 $\pm$ 23 &&& 0.20 $\pm$ 0.08\\ [0.07cm]
&& C1 & 179 $\pm$ 9 & 0.46 $\pm$ 0.08 & $-$119.5 $\pm$ 9.8 & 0.40 $\pm$ 0.16 \\ [0.07cm]
&& E0 & 78 $\pm$ 4 & 7.13 $\pm$ 0.53 & $-$114.8 $\pm$ 4.0 & 1.26 $\pm$ 1.06 \\ [0.07cm]
&& E1 & 117 $\pm$ 6 & 5.30 $\pm$ 0.40 & $-$111.3 $\pm$ 4.0 & 1.22 $\pm$ 0.79 \\ [0.07cm]
&& E4 & 849 $\pm$ 42 & 2.31 $\pm$ 0.07 & $-$114.1 $\pm$ 1.5 & 0.74 $\pm$ 0.13 \\ [0.07cm]
&& E5 & 523 $\pm$ 26 & 1.45 $\pm$ 0.06 & $-$113.3 $\pm$ 2.3 & 0.56 $\pm$ 0.13 \\ [0.07cm]
\hline\\ [0.07cm]
2010.11 & 55238.50 & C0 & 353 $\pm$ 18 &&& 0.16 $\pm$ 0.10\\ [0.07cm]
&& C1 & 204 $\pm$ 10 & 0.46 $\pm$ 0.03 & $-$118.5 $\pm$ 3.9 & 0.20 $\pm$ 0.07 \\ [0.07cm]
&& E0 & 130 $\pm$ 7 & 7.61 $\pm$ 0.60 & $-$115.5 $\pm$ 4.4 & 1.86 $\pm$ 1.20 \\ [0.07cm]
&& E1 & 109 $\pm$ 5 & 5.44 $\pm$ 0.47 & $-$111.3 $\pm$ 3.9 & 1.34 $\pm$ 0.94 \\ [0.07cm]
&& E4 & 602 $\pm$ 30 & 2.79 $\pm$ 0.11 & $-$114.4 $\pm$ 2.1 & 0.93 $\pm$ 0.22 \\ [0.07cm]
&& E5 & 628 $\pm$ 31 & 1.78 $\pm$ 0.05 & $-$113.3 $\pm$ 1.3 & 0.46 $\pm$ 0.09 \\ [0.07cm]
&& S7 & 151 $\pm$ 8 & 1.11 $\pm$ 0.13 & $-$114.8 $\pm$ 6.3 & 0.54 $\pm$ 0.26 \\ [0.07cm]
\hline\\ [0.07cm]
2010.19 & 55265.50 & C0 & 444 $\pm$ 22 &&& 0.24 $\pm$ 0.08\\ [0.07cm]
&& C1 & 136 $\pm$ 7 & 0.53 $\pm$ 0.06 & $-$118.2 $\pm$ 5.8 & 0.25 $\pm$ 0.11 \\ [0.07cm]
&& E0 & 87 $\pm$ 4 & 7.80 $\pm$ 0.50 & $-$116.5 $\pm$ 3.5 & 1.26 $\pm$ 0.99 \\ [0.07cm]
&& E1 & 95 $\pm$ 5 & 6.00 $\pm$ 0.43 & $-$111.2 $\pm$ 4.5 & 1.17 $\pm$ 0.86 \\ [0.07cm]
&& E4 & 392 $\pm$ 20 & 2.97 $\pm$ 0.13 & $-$113.1 $\pm$ 2.4 & 0.90 $\pm$ 0.27 \\ [0.07cm]
&& E5 & 540 $\pm$ 27 & 1.91 $\pm$ 0.05 & $-$112.5 $\pm$ 1.3 & 0.44 $\pm$ 0.09 \\ [0.07cm]
&& S7 & 279 $\pm$ 14 & 1.22 $\pm$ 0.05 & $-$115.9 $\pm$ 2.3 & 0.34 $\pm$ 0.10 \\ [0.07cm]
\hline\\ [0.07cm]
2010.53 & 55389.50 & C0 & 519 $\pm$ 26 &&& 0.24 $\pm$ 0.06 \\ [0.07cm]
&& C1  & 489 $\pm$ 24 & 0.36 $\pm$ 0.03 & $-$130.6 $\pm$ 4.2 & 0.26 $\pm$ 0.05 \\ [0.07cm]
&& E0 & 75 $\pm$ 4 & 8.83 $\pm$ 0.60 & $-$116.4 $\pm$ 4.4 & 1.58 $\pm$ 1.20 \\ [0.07cm]
&& E1 & 77 $\pm$ 4 & 6.89 $\pm$ 0.37 & $-$111.8 $\pm$ 3.8 & 0.92 $\pm$ 0.74 \\ [0.07cm]
&& E4 & 204 $\pm$ 10 & 3.93 $\pm$ 0.18 & $-$114.1 $\pm$ 2.4 & 0.81 $\pm$ 0.35 \\ [0.07cm]
&& E5 & 197 $\pm$ 10 & 2.65 $\pm$ 0.17 & $-$111.1 $\pm$ 3.3 & 0.77 $\pm$ 0.34 \\ [0.07cm]
&& E6 & 284 $\pm$ 14 & 1.81 $\pm$ 0.06 & $-$113.5 $\pm$ 1.8 & 0.39 $\pm$ 0.12 \\ [0.07cm]
&& S7 & 180 $\pm$ 9 & 1.31 $\pm$ 0.08 & $-$117.4 $\pm$ 3.2 & 0.38 $\pm$ 0.15 \\ [0.07cm]
\hline\\ [0.07cm]
2010.65 & 55435.50 & C0 & 345 $\pm$ 17 &&& 0.21 $\pm$ 0.07\\ [0.07cm]
&& C1 & 401 $\pm$ 20 & 0.48 $\pm$ 0.02 & $-$124.9 $\pm$ 2.9 & 0.22 $\pm$ 0.05 \\ [0.07cm]
&& E0 & 45 $\pm$ 2 & 9.83 $\pm$ 0.29 & $-$117.8 $\pm$ 1.6 & 0.58 $\pm$ 0.59 \\ [0.07cm]
&& E1 & 121 $\pm$ 6 & 7.38 $\pm$ 0.60 & $-$113.4 $\pm$ 2.8 & 1.76 $\pm$ 1.20 \\ [0.07cm]
&& E4 & 131 $\pm$ 7 & 4.38 $\pm$ 0.18 & $-$115.5 $\pm$ 2.3 & 0.67 $\pm$ 0.37 \\ [0.07cm]
&& E5 & 174 $\pm$ 9 & 3.28 $\pm$ 0.26 & $-$109.3 $\pm$ 4.1 & 1.04 $\pm$ 0.52 \\ [0.07cm]
&& E6 & 297 $\pm$ 15 & 2.04 $\pm$ 0.09 & $-$114.5 $\pm$ 2.4 & 0.56 $\pm$ 0.18 \\ [0.07cm]
&& S7 & 108 $\pm$ 5 & 1.29 $\pm$ 0.16 & $-$116.2 $\pm$ 6.5 & 0.53 $\pm$ 0.31 \\ [0.07cm]
\hline\\ [0.07cm]
2010.74 & 55468.50 & C0 & 439 $\pm$ 22 &&& 0.23 $\pm$ 0.08\\ [0.07cm]
&& C1 & 297 $\pm$ 15 & 0.60 $\pm$ 0.05 & $-$120.3 $\pm$ 4.3 & 0.32 $\pm$ 0.09 \\ [0.07cm]
&& E0 & 62 $\pm$ 3 & 10.10 $\pm$ 0.25 & $-$118.8 $\pm$ 1.4 & 0.59 $\pm$ 0.50 \\ [0.07cm]
&& E1 & 104 $\pm$ 5 & 7.46 $\pm$ 0.41 & $-$112.7 $\pm$ 4.4 & 1.17 $\pm$ 0.82 \\ [0.07cm]
&& E4 & 122 $\pm$ 6 & 4.54 $\pm$ 0.18 & $-$116.3 $\pm$ 2.1 & 0.64 $\pm$ 0.36 \\ [0.07cm]
&& E5 & 160 $\pm$ 8 & 3.62 $\pm$ 0.20 & $-$108.0 $\pm$ 2.8 & 0.79 $\pm$ 0.39 \\ [0.07cm]
&& E6 & 235 $\pm$ 12 & 2.21 $\pm$ 0.12 & $-$112.3 $\pm$ 2.9 & 0.64 $\pm$ 0.24 \\ [0.07cm]
&& S7 & 131 $\pm$ 7 & 1.34 $\pm$ 0.11 & $-$112.6 $\pm$ 4.4 & 0.45 $\pm$ 0.23 \\ [0.07cm]
\hline\\ [0.07cm]
2010.84 & 55504.50 & C0 & 586 $\pm$ 29 &&& 0.24 $\pm$ 0.07\\ [0.07cm]
&& E0 & 63 $\pm$ 3 & 10.39 $\pm$ 0.17 & $-$118.9 $\pm$ 0.9 & 0.44 $\pm$ 0.34 \\ [0.07cm]
&& E1 & 103 $\pm$ 5 & 8.07 $\pm$ 0.60 & $-$114.2 $\pm$ 2.9 & 1.93 $\pm$ 1.20 \\ [0.07cm]
&& E4 & 152 $\pm$ 8 & 4.74 $\pm$ 0.20 & $-$115.2 $\pm$ 2.3 & 0.78 $\pm$ 0.40 \\ [0.07cm]
&& E5 & 116 $\pm$ 6 & 3.85 $\pm$ 0.16 & $-$107.7 $\pm$ 2.1 & 0.56 $\pm$ 0.32 \\ [0.07cm]
&& E6 & 208 $\pm$ 10 & 2.41 $\pm$ 0.15 & $-$111.9 $\pm$ 3.3 & 0.72 $\pm$ 0.30 \\ [0.07cm]
&& E8 & 231 $\pm$ 12 & 0.75 $\pm$ 0.10 & $-$120.7 $\pm$ 4.1 & 0.53 $\pm$ 0.19 \\ [0.07cm]
&& S7 & 170 $\pm$ 9 & 1.48 $\pm$ 0.09 & $-$116.2 $\pm$ 3.5 & 0.44 $\pm$ 0.19 \\ [0.07cm]
\hline\\ [0.07cm]
2010.91 & 55529.50 & C0 & 529 $\pm$ 26 &&& 0.20 $\pm$ 0.06\\ [0.07cm]
&& C1 & 104 $\pm$ 5 & 0.39 $\pm$ 0.05 & $-$128.8 $\pm$ 7.9 & 0.21 $\pm$ 0.11 \\ [0.07cm]
&& E0 & 62 $\pm$ 3 & 10.60 $\pm$ 0.23 & $-$118.7 $\pm$ 1.2 & 0.57 $\pm$ 0.47 \\ [0.07cm]
&& E1 & 105 $\pm$ 5 & 8.23 $\pm$ 0.60 & $-$113.7 $\pm$ 5.0 & 1.93 $\pm$ 1.20 \\ [0.07cm]
&& E4 & 138 $\pm$ 7 & 4.88 $\pm$ 0.24 & $-$115.8 $\pm$ 2.6 & 0.85 $\pm$ 0.47 \\ [0.07cm]
&& E5 & 97 $\pm$ 5 & 4.01 $\pm$ 0.18 & $-$107.5 $\pm$ 2.3 & 0.57 $\pm$ 0.36 \\ [0.07cm]
&& E6 & 139 $\pm$ 7 & 2.62 $\pm$ 0.21 & $-$111.6 $\pm$ 4.2 & 0.77 $\pm$ 0.41 \\ [0.07cm]
&& E8 & 168 $\pm$ 8 & 1.07 $\pm$ 0.09 & $-$118.2 $\pm$ 4.4 & 0.41 $\pm$ 0.18 \\ [0.07cm]
&& S7 & 154 $\pm$ 8 & 1.61 $\pm$ 0.08 & $-$115.2 $\pm$ 3.4 & 0.37 $\pm$ 0.16 \\ [0.07cm]
\hline\\ [0.07cm]
2010.98 & 55554.50 & C0 & 563 $\pm$ 28 &&& 0.17 $\pm$ 0.05\\ [0.07cm]
&& C1 & 231 $\pm$ 12 & 0.34 $\pm$ 0.03 & $-$123.3 $\pm$ 4.3 & 0.18 $\pm$ 0.05 \\ [0.07cm]
&& E0 & 38 $\pm$ 2 & 10.64 $\pm$ 0.18 & $-$116.9 $\pm$ 1.0 & 0.35 $\pm$ 0.36 \\ [0.07cm]
&& E1 & 102 $\pm$ 5 & 8.41 $\pm$ 0.48 & $-$114.5 $\pm$ 4.8 & 1.33 $\pm$ 0.96 \\ [0.07cm]
&& E4 & 132 $\pm$ 7 & 5.11 $\pm$ 0.20 & $-$116.9 $\pm$ 2.1 & 0.71 $\pm$ 0.39 \\ [0.07cm]
&& E5 & 132 $\pm$ 7 & 4.24 $\pm$ 0.21 & $-$108.6 $\pm$ 2.5 & 0.76 $\pm$ 0.42 \\ [0.07cm]
&& E6 & 123 $\pm$ 6 & 2.84 $\pm$ 0.24 & $-$112.7 $\pm$ 4.4 & 0.81 $\pm$ 0.48 \\ [0.07cm]
&& E8 & 331 $\pm$ 17 & 1.29 $\pm$ 0.06 & $-$117.7 $\pm$ 2.7 & 0.44 $\pm$ 0.13 \\ [0.07cm]
&& S7 & 100 $\pm$ 5 & 1.82 $\pm$ 0.05 & $-$116.3 $\pm$ 3.2 & 0.19 $\pm$ 0.09 \\ [0.07cm]
\hline\\ [0.07cm]
2011.14 & 55612.50 & C0 & 814 $\pm$ 41 &&& 0.18 $\pm$ 0.05\\ [0.07cm]
&& C1 & 187 $\pm$ 9 & 0.42 $\pm$ 0.05 & $-$117.9 $\pm$ 7.2 & 0.29 $\pm$ 0.11 \\ [0.07cm]
&& E0 & 65 $\pm$ 3 & 11.21 $\pm$ 0.23 & $-$118.9 $\pm$ 1.1 & 0.56 $\pm$ 0.45 \\ [0.07cm]
&& E1 & 102 $\pm$ 5 & 8.94 $\pm$ 0.60 & $-$114.7 $\pm$ 3.1 & 1.90 $\pm$ 1.20 \\ [0.07cm]
&& E4 & 145 $\pm$ 7 & 5.38 $\pm$ 0.27 & $-$114.5 $\pm$ 2.7 & 0.98 $\pm$ 0.54 \\ [0.07cm]
&& E5 & 53 $\pm$ 3 & 4.50 $\pm$ 0.25 & $-$107.5 $\pm$ 2.9 & 0.56 $\pm$ 0.51 \\ [0.07cm]
&& E6 & 101 $\pm$ 5 & 3.36 $\pm$ 0.30 & $-$107.6 $\pm$ 4.6 & 0.90 $\pm$ 0.60 \\ [0.07cm]
&& E8 & 410 $\pm$ 20 & 1.54 $\pm$ 0.06 & $-$115.2 $\pm$ 2.1 & 0.48 $\pm$ 0.12 \\ [0.07cm]
&& S7 & 93 $\pm$ 5 & 2.08 $\pm$ 0.13 & $-$115.6 $\pm$ 3.2 & 0.41 $\pm$ 0.25 \\ [0.07cm]
\hline\\ [0.07cm]
2011.16 & 55619.50 & C0 & 1017 $\pm$ 51 &&& 0.18 $\pm$ 0.05\\ [0.07cm]
&& C1 & 225 $\pm$ 11 & 0.40 $\pm$ 0.05 & $-$123.2 $\pm$ 6.9 & 0.30 $\pm$ 0.10 \\ [0.07cm]
&& E0 & 76 $\pm$ 4 & 11.27 $\pm$ 0.21 & $-$118.9 $\pm$ 1.0 & 0.59 $\pm$ 0.43 \\ [0.07cm]
&& E1 & 121 $\pm$ 6 & 8.93 $\pm$ 0.60 & $-$114.7 $\pm$ 4.4 & 1.85 $\pm$ 1.20 \\ [0.07cm]
&& E4 & 178 $\pm$ 9 & 5.41 $\pm$ 0.25 & $-$113.8 $\pm$ 2.5 & 1.01 $\pm$ 0.50 \\ [0.07cm]
&& E5 & 52 $\pm$ 3 & 4.52 $\pm$ 0.26 & $-$108.2 $\pm$ 2.9 & 0.57 $\pm$ 0.52 \\ [0.07cm]
&& E6 & 112 $\pm$ 6 & 3.43 $\pm$ 0.24 & $-$108.3 $\pm$ 3.5 & 0.77 $\pm$ 0.47 \\ [0.07cm]
&& E8 & 463 $\pm$ 23 & 1.57 $\pm$ 0.06 & $-$115.3 $\pm$ 2.0 & 0.49 $\pm$ 0.12 \\ [0.07cm]
\hline\\ [0.07cm]
2011.17 & 55625.50 & C0 & 954 $\pm$ 48 &&& 0.16 $\pm$ 0.06 \\ [0.07cm]
&& C1 & 386 $\pm$ 19 & 0.32 $\pm$ 0.02 & $-$124.8 $\pm$ 4.0 & 0.20 $\pm$ 0.05 \\ [0.07cm]
&& E0 & 80 $\pm$ 4 & 11.29 $\pm$ 0.21 & $-$118.4 $\pm$ 1.0 & 0.59 $\pm$ 0.42 \\ [0.07cm]
&& E1 & 107 $\pm$ 5 & 9.02 $\pm$ 0.60 & $-$114.6 $\pm$ 3.9 & 1.97 $\pm$ 1.20 \\ [0.07cm]
&& E4 & 186 $\pm$ 9 & 5.45 $\pm$ 0.28 & $-$113.8 $\pm$ 2.7 & 1.13 $\pm$ 0.55 \\ [0.07cm]
&& E5 & 44 $\pm$ 2 & 4.68 $\pm$ 0.14 & $-$108.8 $\pm$ 1.5 & 0.30 $\pm$ 0.27 \\ [0.07cm]
&& E6 & 162 $\pm$ 8 & 3.35 $\pm$ 0.32 & $-$108.9 $\pm$ 4.9 & 1.19 $\pm$ 0.64 \\ [0.07cm]
&& E8 & 385 $\pm$ 19 & 1.78 $\pm$ 0.05 & $-$114.3 $\pm$ 1.5 & 0.39 $\pm$ 0.10 \\ [0.07cm]
&& S8 & 199 $\pm$ 10 & 1.31 $\pm$ 0.08 & $-$118.1 $\pm$ 3.2 & 0.40 $\pm$ 0.15 \\ [0.07cm]
\hline\\ [0.07cm]
2011.38 & 55702.50 & C0 & 689 $\pm$ 34 &&& 0.17 $\pm$ 0.09 \\ [0.07cm]
&& C1 & 343 $\pm$ 17 & 0.49 $\pm$ 0.05 & $-$116.7 $\pm$ 5.1 & 0.34 $\pm$ 0.09 \\ [0.07cm]
&& E0 & 64 $\pm$ 3 & 11.70 $\pm$ 0.29 & $-$117.9 $\pm$ 1.4 & 0.69 $\pm$ 0.58 \\ [0.07cm]
&& E1 & 61 $\pm$ 3 & 9.51 $\pm$ 0.60 & $-$116.5 $\pm$ 4.5 & 1.37 $\pm$ 1.20 \\ [0.07cm]
&& E4 & 85 $\pm$ 5 & 6.02 $\pm$ 0.35 & $-$112.4 $\pm$ 3.1 & 1.06 $\pm$ 0.70 \\ [0.07cm]
&& E5 & 0.08 $\pm$ 4 & 5.09 $\pm$ 0.42 & $-$114.2 $\pm$ 4.4 & 1.08 $\pm$ 0.84 \\ [0.07cm]
&& E6 & 110 $\pm$ 6 & 3.75 $\pm$ 0.20 & $-$106.3 $\pm$ 2.7 & 0.67 $\pm$ 0.41 \\ [0.07cm]
&& E8 & 301 $\pm$ 15 & 2.08 $\pm$ 0.06 & $-$112.8 $\pm$ 1.5 & 0.39 $\pm$ 0.12 \\ [0.07cm]
&& S8 & 93 $\pm$ 5 & 1.41 $\pm$ 0.10 & $-$120.4 $\pm$ 3.9 & 0.34 $\pm$ 0.20 \\ [0.07cm]
\hline\\ [0.07cm]
2011.56 & 55766.50 & C0 & 483 $\pm$ 24 &&& 0.24 $\pm$ 0.10\\ [0.07cm]
&& E0 & 59 $\pm$ 3 & 11.96 $\pm$ 0.34 & $-$117.6 $\pm$ 1.5 & 0.75 $\pm$ 0.67 \\ [0.07cm]
&& E1 & 68 $\pm$ 3 & 9.94 $\pm$ 0.60 & $-$115.5 $\pm$ 3.9 & 1.93 $\pm$ 1.20 \\ [0.07cm]
&& E4 & 86 $\pm$ 4 & 6.54 $\pm$ 0.38 & $-$112.3 $\pm$ 3.1 & 1.01 $\pm$ 0.77 \\ [0.07cm]
&& E5 & 67 $\pm$ 3 & 5.50 $\pm$ 0.39 & $-$113.6 $\pm$ 3.8 & 0.91 $\pm$ 0.79 \\ [0.07cm]
&& E6 & 144 $\pm$ 7 & 4.05 $\pm$ 0.26 & $-$109.1 $\pm$ 3.3 & 0.93 $\pm$ 0.51 \\ [0.07cm]
&& E8 & 162 $\pm$ 8 & 2.40 $\pm$ 0.08 & $-$112.3 $\pm$ 1.7 & 0.36 $\pm$ 0.15 \\ [0.07cm]
&& E9 & 214 $\pm$ 11 & 0.71 $\pm$ 0.09 & $-$117.8 $\pm$ 4.0 & 0.49 $\pm$ 0.19 \\ [0.07cm]
&& S8 & 88 $\pm$ 4 & 1.71 $\pm$ 0.18 & $-$115.9 $\pm$ 5.6 & 0.54 $\pm$ 0.35 \\ [0.07cm]
\hline\\ [0.07cm]
2011.65 & 55799.50 & C0 & 505 $\pm$ 25 &&& 0.12 $\pm$ 0.08\\ [0.07cm]
&& C1 & 136 $\pm$ 7 & 0.45 $\pm$ 0.07 & $-$121.9 $\pm$ 9.1 & 0.32 $\pm$ 0.15 \\ [0.07cm]
&& E0 & 55 $\pm$ 3 & 12.26 $\pm$ 0.39 & $-$117.7 $\pm$ 1.7 & 0.82 $\pm$ 0.78 \\ [0.07cm]
&& E1 & 68 $\pm$ 3 & 10.38 $\pm$ 0.60 & $-$115.4 $\pm$ 5.2 & 1.95 $\pm$ 1.20 \\ [0.07cm]
&& E4 & 75 $\pm$ 4 & 6.92 $\pm$ 0.42 & $-$112.3 $\pm$ 3.2 & 1.01 $\pm$ 0.83 \\ [0.07cm]
&& E5 & 73 $\pm$ 4 & 5.90 $\pm$ 0.41 & $-$113.8 $\pm$ 3.7 & 0.98 $\pm$ 0.82 \\ [0.07cm]
&& E6 & 138 $\pm$ 7 & 4.34 $\pm$ 0.25 & $-$110.8 $\pm$ 3.0 & 0.89 $\pm$ 0.49 \\ [0.07cm]
&& E8 & 190 $\pm$ 10 & 2.55 $\pm$ 0.15 & $-$112.9 $\pm$ 3.2 & 0.70 $\pm$ 0.31 \\ [0.07cm]
&& E9 & 161 $\pm$ 8 & 1.18 $\pm$ 0.13 & $-$116.5 $\pm$ 4.0 & 0.57 $\pm$ 0.27 \\ [0.07cm]
\hline\\ [0.07cm]
2011.75 & 55837.50 & C0 & 625 $\pm$ 31 &&& 0.14 $\pm$ 0.08\\ [0.07cm]
&& C1 & 146 $\pm$ 7 & 0.37 $\pm$ 0.06 & $-$129.8 $\pm$ 9.1 & 0.27 $\pm$ 0.12 \\ [0.07cm]
&& E0 & 47 $\pm$ 2 & 12.49 $\pm$ 0.45 & $-$117.8 $\pm$ 1.9 & 0.86 $\pm$ 0.91 \\ [0.07cm]
&& E1 & 72 $\pm$ 4 & 10.62 $\pm$ 0.60 & $-$115.4 $\pm$ 5.1 & 2.19 $\pm$ 1.20 \\ [0.07cm]
&& E4 & 69 $\pm$ 3 & 7.14 $\pm$ 0.46 & $-$111.5 $\pm$ 3.4 & 1.06 $\pm$ 0.92 \\ [0.07cm]
&& E5 & 63 $\pm$ 3 & 6.13 $\pm$ 0.36 & $-$115.0 $\pm$ 3.2 & 0.82 $\pm$ 0.72 \\ [0.07cm]
&& E6 & 158 $\pm$ 8 & 4.57 $\pm$ 0.26 & $-$111.0 $\pm$ 3.0 & 0.99 $\pm$ 0.52 \\ [0.07cm]
&& E8 & 108 $\pm$ 5 & 2.82 $\pm$ 0.11 & $-$111.3 $\pm$ 2.0 & 0.39 $\pm$ 0.21 \\ [0.07cm]
&& E9 & 195 $\pm$ 10 & 1.27 $\pm$ 0.08 & $-$117.8 $\pm$ 3.3 & 0.39 $\pm$ 0.15 \\ [0.07cm]
\hline\\ [0.07cm]
2011.94 & 55907.50 & C0 & 420 $\pm$ 21 &&& 0.17 $\pm$ 0.08\\ [0.07cm]
&& C1 & 143 $\pm$ 7 & 0.45 $\pm$ 0.06 & $-$124.8 $\pm$ 7.9 & 0.29 $\pm$ 0.13 \\ [0.07cm]
&& E0 & 38 $\pm$ 2 & 12.78 $\pm$ 0.60 & $-$117.8 $\pm$ 3.3 & 1.20 $\pm$ 1.20 \\ [0.07cm]
&& E1 & 55 $\pm$ 3 & 10.71 $\pm$ 0.60 & $-$114.7 $\pm$ 5.5 & 2.28 $\pm$ 1.20 \\ [0.07cm]
&& E4 & 82 $\pm$ 4 & 7.11 $\pm$ 0.49 & $-$113.3 $\pm$ 3.7 & 1.21 $\pm$ 0.98 \\ [0.07cm]
&& E6 & 140 $\pm$ 7 & 5.13 $\pm$ 0.30 & $-$112.1 $\pm$ 3.1 & 1.04 $\pm$ 0.60 \\ [0.07cm]
&& E8 & 88 $\pm$ 4 & 3.21 $\pm$ 0.20 & $-$110.6 $\pm$ 3.2 & 0.59 $\pm$ 0.39 \\ [0.07cm]
&& E9 & 256 $\pm$ 13 & 1.65 $\pm$ 0.10 & $-$116.3 $\pm$ 3.3 & 0.57 $\pm$ 0.20 \\ [0.07cm]
\hline\\ [0.07cm]
2011.99 & 55924.50 & C0 & 444 $\pm$ 22 &&& 0.16 $\pm$ 0.07\\ [0.07cm]
&& C1 & 170 $\pm$ 8 & 0.35 $\pm$ 0.06 & $-$109.7 $\pm$ 9.2 & 0.31 $\pm$ 0.13 \\ [0.07cm]
&& E0 & 36 $\pm$ 2 & 13.03 $\pm$ 0.60 & $-$117.4 $\pm$ 3.4 & 1.22 $\pm$ 1.20 \\ [0.07cm]
&& E1 & 55 $\pm$ 3 & 11.15 $\pm$ 0.60 & $-$115.5 $\pm$ 6.7 & 1.72 $\pm$ 1.20 \\ [0.07cm]
&& E4 & 88 $\pm$ 4 & 7.17 $\pm$ 0.53 & $-$112.9 $\pm$ 3.9 & 1.33 $\pm$ 1.06 \\ [0.07cm]
&& E6 & 146 $\pm$ 7 & 5.23 $\pm$ 0.30 & $-$111.7 $\pm$ 7.1 & 1.08 $\pm$ 0.61 \\ [0.07cm]
&& E8 & 101 $\pm$ 5 & 3.32 $\pm$ 0.20 & $-$110.1 $\pm$ 3.2 & 0.64 $\pm$ 0.40 \\ [0.07cm]
&& E9 & 154 $\pm$ 13 & 1.71 $\pm$ 0.11 & $-$115.2 $\pm$ 3.5 & 0.61 $\pm$ 0.22 \\ [0.07cm]
\hline\\ [0.07cm]
2012.04 & 55940.50 & C0 & 494 $\pm$ 25 &&& 0.16 $\pm$ 0.07\\ [0.07cm]
&& C1 & 144 $\pm$ 7 & 0.48 $\pm$ 0.08 & $-$121.6 $\pm$ 9.3 & 0.35 $\pm$ 0.16 \\ [0.07cm]
&& E0 & 38 $\pm$ 1 & 12.95 $\pm$ 0.60 & $-$117.4 $\pm$ 3.8 & 1.37 $\pm$ 1.20 \\ [0.07cm]
&& E1 & 49 $\pm$ 2 & 11.17 $\pm$ 0.60 & $-$115.1 $\pm$ 4.6 & 1.94 $\pm$ 1.20 \\ [0.07cm]
&& E4 & 87 $\pm$ 4 & 7.31 $\pm$ 0.60 & $-$113.1 $\pm$ 4.4 & 1.49 $\pm$ 1.20 \\ [0.07cm]
&& E6 & 135 $\pm$ 7 & 5.39 $\pm$ 0.28 & $-$111.9 $\pm$ 3.1 & 0.97 $\pm$ 0.56 \\ [0.07cm]
&& E8 & 116 $\pm$ 6 & 3.44 $\pm$ 0.19 & $-$111.1 $\pm$ 2.9 & 0.66 $\pm$ 0.39 \\ [0.07cm]
&& E9 & 228 $\pm$ 11 & 1.80 $\pm$ 0.13 & $-$115.6 $\pm$ 4.0 & 0.68 $\pm$ 0.26 \\ [0.07cm]
\hline\\ [0.07cm]
2012.17 & 55990.50 & C0 & 559 $\pm$ 28 &&& 0.16 $\pm$ 0.06\\ [0.07cm]
&& C1 & 111 $\pm$ 6 & 0.70 $\pm$ 0.14 & $-$121.6 $\pm$ 28.6 & 0.50 $\pm$ 0.29 \\ [0.07cm]
&& E0 & 24 $\pm$ 2 & 12.97 $\pm$ 0.41 & $-$118.3 $\pm$ 1.7 & 0.56 $\pm$ 0.82 \\ [0.07cm]
&& E1 & 48 $\pm$ 2 & 11.89 $\pm$ 0.60 & $-$114.6 $\pm$ 5.7 & 2.25 $\pm$ 1.20 \\ [0.07cm]
&& E4 & 70 $\pm$ 4 & 7.75 $\pm$ 0.60 & $-$112.8 $\pm$ 4.9 & 1.53 $\pm$ 1.20 \\ [0.07cm]
&& E6 & 128 $\pm$ 6 & 5.77 $\pm$ 0.31 & $-$111.9 $\pm$ 2.7 & 1.03 $\pm$ 0.62 \\ [0.07cm]
&& E8 & 199 $\pm$ 10 & 3.69 $\pm$ 0.11 & $-$110.2 $\pm$ 1.5 & 0.53 $\pm$ 0.21 \\ [0.07cm]
&& E9 & 145 $\pm$ 7 & 2.02 $\pm$ 0.19 & $-$116.1 $\pm$ 5.2 & 0.74 $\pm$ 0.39 \\ [0.07cm]
\hline\\ [0.07cm]
2012.33 & 56046.50 & C0 & 560 $\pm$ 28 &&& 0.17 $\pm$ 0.06\\ [0.07cm]
&& C1 & 140 $\pm$ 7 & 0.37 $\pm$ 0.06 & $-$123.9 $\pm$ 26.1 & 0.29 $\pm$ 0.13 \\ [0.07cm]
&& E1 & 75 $\pm$ 4 & 12.84 $\pm$ 0.60 & $-$116.1 $\pm$ 6.4 & 1.93 $\pm$ 1.20 \\ [0.07cm]
&& E4 & 66 $\pm$ 3 & 8.11 $\pm$ 0.60 & $-$113.4 $\pm$ 5.0 & 1.58 $\pm$ 1.20 \\ [0.07cm]
&& E6 & 123 $\pm$ 6 & 6.25 $\pm$ 0.33 & $-$112.4 $\pm$ 3.8 & 1.06 $\pm$ 0.66 \\ [0.07cm]
&& E8 & 279 $\pm$ 14 & 3.98 $\pm$ 0.09 & $-$110.8 $\pm$ 1.2 & 0.54 $\pm$ 0.18 \\ [0.07cm]
&& E9 & 94 $\pm$ 5 & 2.47 $\pm$ 0.27 & $-$115.1 $\pm$ 5.9 & 0.79 $\pm$ 0.54 \\ [0.07cm]
&& S9 & 68 $\pm$ 3 & 1.36 $\pm$ 0.28 & $-$120.0 $\pm$ 8.0 & 0.68 $\pm$ 0.55 \\ [0.07cm]
\hline\\ [0.07cm]
2012.39 & 56071.50 & C0 & 509 $\pm$ 25 &&& 0.13 $\pm$ 0.05\\ [0.07cm]
&& C1 & 162 $\pm$ 8 & 0.32 $\pm$ 0.05 & $-$125.6 $\pm$ 8.0 & 0.24 $\pm$ 0.09 \\ [0.07cm]
&& E1 & 63 $\pm$ 3 & 12.96 $\pm$ 0.60 & $-$116.4 $\pm$ 3.8 & 1.67 $\pm$ 1.20 \\ [0.07cm]
&& E4 & 42 $\pm$ 2 & 8.62 $\pm$ 0.60 & $-$113.4 $\pm$ 5.1 & 1.34 $\pm$ 1.20 \\ [0.07cm]
&& E6 & 138 $\pm$ 7 & 6.42 $\pm$ 0.53 & $-$112.4 $\pm$ 2.8 & 1.67 $\pm$ 1.05 \\ [0.07cm]
&& E8 & 289 $\pm$ 14 & 4.14 $\pm$ 0.08 & $-$111.1 $\pm$ 1.0 & 0.51 $\pm$ 0.16 \\ [0.07cm]
&& E9 & 80 $\pm$ 4 & 2.71 $\pm$ 0.21 & $-$115.9 $\pm$ 4.2 & 0.59 $\pm$ 0.42 \\ [0.07cm]
&& S9 & 73 $\pm$ 4 & 1.44 $\pm$ 0.30 & $-$118.3 $\pm$ 8.0 & 0.75 $\pm$ 0.59 \\ [0.07cm]
\hline\\ [0.07cm]
2012.53 & 56120.50 & C0 & 548 $\pm$ 27 &&& 0.15 $\pm$ 0.06\\ [0.07cm]
&& C1 & 166 $\pm$ 8 & 0.27 $\pm$ 0.02 & $-$118.0 $\pm$ 4.6 & 0.13 $\pm$ 0.05 \\ [0.07cm]
&& C2 & 50 $\pm$ 3 & 0.65 $\pm$ 0.13 & $-$120.2 $\pm$ 6.0 & 0.31 $\pm$ 0.26 \\ [0.07cm]
&& E1 & 57 $\pm$ 3 & 13.31 $\pm$ 0.60 & $-$116.3 $\pm$ 3.5 & 1.78 $\pm$ 1.20 \\ [0.07cm]
&& E4 & 47 $\pm$ 2 & 8.68 $\pm$ 0.60 & $-$113.5 $\pm$ 5.1 & 1.40 $\pm$ 1.20 \\ [0.07cm]
&& E6 & 99 $\pm$ 5 & 6.78 $\pm$ 0.42 & $-$112.4 $\pm$ 4.3 & 1.17 $\pm$ 0.84 \\ [0.07cm]
&& E8 & 263 $\pm$ 13 & 4.46 $\pm$ 0.13 & $-$111.8 $\pm$ 1.5 & 0.70 $\pm$ 0.25 \\ [0.07cm]
&& E9 & 60 $\pm$ 3 & 2.99 $\pm$ 0.50 & $-$113.1 $\pm$ 7.0 & 1.05 $\pm$ 0.10 \\ [0.07cm]
&& S9 & 45 $\pm$ 2 & 1.68 $\pm$ 0.52 & $-$118.7 $\pm$ 9.5 & 0.95 $\pm$ 1.04 \\ [0.07cm]
\hline\\ [0.07cm]
2012.59 & 56142.50 & C0 & 490 $\pm$ 25 &&& $<$ 0.05\\ [0.07cm]
&& C1 & 319 $\pm$ 16 & 0.23 $\pm$ 0.01 & $-$115.0 $\pm$ 2.5 & 0.09 $\pm$ 0.02 \\ [0.07cm]
&& C2 & 49 $\pm$ 2 & 0.65 $\pm$ 0.04 & $-$116.9 $\pm$ 3.1 & 0.11 $\pm$ 0.07 \\ [0.07cm]
&& E1 & 50 $\pm$ 2 & 13.55 $\pm$ 0.60 & $-$115.7 $\pm$ 3.9 & 1.65 $\pm$ 1.20 \\ [0.07cm]
&& E4 & 43 $\pm$ 2 & 8.90 $\pm$ 0.60 & $-$113.5 $\pm$ 5.9 & 1.56 $\pm$ 1.20 \\ [0.07cm]
&& E6 & 93 $\pm$ 5 & 6.99 $\pm$ 0.42 & $-$112.5 $\pm$ 3.3 & 1.13 $\pm$ 0.83 \\ [0.07cm]
&& E8 & 238 $\pm$ 12 & 4.66 $\pm$ 0.13 & $-$111.9 $\pm$ 1.5 & 0.70 $\pm$ 0.27 \\ [0.07cm]
&& E9 & 69 $\pm$ 3 & 3.34 $\pm$ 0.57 & $-$112.5 $\pm$ 7.0 & 1.26 $\pm$ 1.15 \\ [0.07cm]
&& S9 & 46 $\pm$ 2 & 1.55 $\pm$ 0.56 & $-$118.7 $\pm$ 9.5 & 1.02 $\pm$ 1.13 \\ [0.07cm]
\hline\\ [0.07cm]
2012.67 & 56172.50 & C0 & 582 $\pm$ 29 &&& $<$ 0.05 \\ [0.07cm]
&& C1 & 424 $\pm$ 21 & 0.22 $\pm$ 0.01 & $-$117.5 $\pm$ 1.6 & 0.08 $\pm$ 0.01 \\ [0.07cm]
&& C2 & 63 $\pm$ 3 & 0.64 $\pm$ 0.06 & $-$115.9 $\pm$ 5.5 & 0.20 $\pm$ 0.13 \\ [0.07cm]
&& E1 & 65 $\pm$ 3 & 13.69 $\pm$ 0.60 & $-$115.5 $\pm$ 3.8 & 2.29 $\pm$ 1.20 \\ [0.07cm]
&& E4 & 40 $\pm$ 2 & 9.35 $\pm$ 0.60 & $-$113.8 $\pm$ 6.0 & 1.60 $\pm$ 1.20 \\ [0.07cm]
&& E6 & 100 $\pm$ 5 & 7.24 $\pm$ 0.50 & $-$112.5 $\pm$ 3.2 & 1.35 $\pm$ 0.99 \\ [0.07cm]
&& E8 & 208 $\pm$ 10 & 4.91 $\pm$ 0.17 & $-$112.1 $\pm$ 1.8 & 0.79 $\pm$ 0.34 \\ [0.07cm]
&& E9 & 68 $\pm$ 3 & 3.59 $\pm$ 0.58 & $-$112.4 $\pm$ 7.0 & 1.27 $\pm$ 1.16 \\ [0.07cm]
&& S9 & 54 $\pm$ 3 & 1.53 $\pm$ 0.40 & $-$120.3 $\pm$ 9.6 & 0.83 $\pm$ 0.81 \\ [0.07cm]
\hline\\ [0.07cm]
2012.84 & 56233.50 & C0 & 641 $\pm$ 32 &&& 0.10 $\pm$ 0.07\\ [0.07cm]
&& C1 & 478 $\pm$ 24 & 0.23 $\pm$ 0.01 & $-$113.1 $\pm$ 2.2 & 0.11 $\pm$ 0.019 \\ [0.07cm]
&& C2 & 70 $\pm$ 3 & 0.59 $\pm$ 0.05 & $-$113.2 $\pm$ 4.1 & 0.16 $\pm$ 0.09 \\ [0.07cm]
&& E1 & 57 $\pm$ 3 & 14.17 $\pm$ 0.60 & $-$115.0 $\pm$ 4.8 & 2.30 $\pm$ 1.20 \\ [0.07cm]
&& E4 & 34 $\pm$ 2 & 9.76 $\pm$ 0.60 & $-$113.6 $\pm$ 6.2 & 1.56 $\pm$ 1.20 \\ [0.07cm]
&& E6 & 74 $\pm$ 4 & 7.70 $\pm$ 0.60 & $-$112.2 $\pm$ 3.6 & 1.40 $\pm$ 1.20 \\ [0.07cm]
&& E8 & 181 $\pm$ 9 & 5.32 $\pm$ 0.22 & $-$112.5 $\pm$ 2.2 & 0.92 $\pm$ 0.44 \\ [0.07cm]
&& E9 & 56 $\pm$ 3 & 3.87 $\pm$ 0.57 & $-$112.9 $\pm$ 7.0 & 1.14 $\pm$ 1.15 \\ [0.07cm]
&& S9 & 43 $\pm$ 2 & 1.44 $\pm$ 0.48 & $-$119.6 $\pm$ 9.7 & 0.87 $\pm$ 0.97 \\ [0.07cm]
\hline\\ [0.07cm]
2012.91 & 56259.50 & C0 & 491 $\pm$ 25 &&& $<$ 0.05 \\ [0.07cm]
&& C1 & 590 $\pm$ 29 & 0.23 $\pm$ 0.01 & $-$116.5 $\pm$ 0.6 & 0.05 $\pm$ 0.05 \\ [0.07cm]
&& C2 & 123 $\pm$ 6 & 0.55 $\pm$ 0.03 & $-$114.8 $\pm$ 3.2 & 0.16 $\pm$ 0.07 \\ [0.07cm]
&& E1 & 40 $\pm$ 2 & 14.29 $\pm$ 0.60 & $-$115.7 $\pm$ 4.9 & 1.57 $\pm$ 1.20 \\ [0.07cm]
&& E4 & 26 $\pm$ 1 & 9.97 $\pm$ 0.60 & $-$113.6 $\pm$ 4.6 & 1.09 $\pm$ 1.20 \\ [0.07cm]
&& E6 & 64 $\pm$ 3 & 8.01 $\pm$ 0.60 & $-$112.3 $\pm$ 4.3 & 1.42 $\pm$ 1.20 \\ [0.07cm]
&& E8 & 178 $\pm$ 9 & 5.53 $\pm$ 0.24 & $-$112.8 $\pm$ 2.3 & 0.99 $\pm$ 0.49 \\ [0.07cm]
&& E9 & 52 $\pm$ 3 & 4.01 $\pm$ 0.54 & $-$113.9 $\pm$ 5.7 & 1.04 $\pm$ 1.07 \\ [0.07cm]
&& S9 & 43 $\pm$ 2 & 1.31 $\pm$ 0.43 & $-$116.4 $\pm$ 9.0 & 0.79 $\pm$ 0.87 \\ [0.07cm]
\hline\\ [0.07cm]
2012.98 & 56284.50 & C0 & 633 $\pm$ 32 &&& 0.06 $\pm$ 0.05 \\ [0.07cm]
&& C1 & 535 $\pm$ 27 & 0.26 $\pm$ 0.01 & $-$118.0 $\pm$ 2.0 & $<$ 0.05 \\ [0.07cm]
&& C2 & 82 $\pm$ 4 & 0.70 $\pm$ 0.01 & $-$116.7 $\pm$ 3.2 & $<$ 0.05 \\ [0.07cm]
&& E1 & 45 $\pm$ 2 & 14.51 $\pm$ 0.60 & $-$114.9 $\pm$ 3.9 & 1.85 $\pm$ 1.20 \\ [0.07cm]
&& E4 & 33 $\pm$ 2 & 10.52 $\pm$ 0.60 & $-$113.4 $\pm$ 5.8 & 1.56 $\pm$ 1.20 \\ [0.07cm]
&& E6 & 66 $\pm$ 3 & 8.01 $\pm$ 0.60 & $-$112.7 $\pm$ 4.6 & 1.39 $\pm$ 1.20 \\ [0.07cm]
&& E8 & 176 $\pm$ 9 & 5.63 $\pm$ 0.25 & $-$112.8 $\pm$ 2.3 & 1.01 $\pm$ 0.49 \\ [0.07cm]
&& E9 & 61 $\pm$ 3 & 4.11 $\pm$ 0.44 & $-$113.7 $\pm$ 5.9 & 0.95 $\pm$ 0.88 \\ [0.07cm]
&& S9 & 27 $\pm$ 1 & 1.59 $\pm$ 0.24 & $-$115.1 $\pm$ 8.1 & 0.38 $\pm$ 0.48 \\ [0.07cm]
\hline\\ [0.07cm]
2013.05 & 56313.50 & C0 & 665 $\pm$ 33 &&& $<$ 0.05 \\ [0.07cm]
&& C1 & 575 $\pm$ 29 & 0.23 $\pm$ 0.01 & $-$117.6 $\pm$ 2.5 & 0.08 $\pm$ 0.05 \\ [0.07cm]
&& C2 & 109 $\pm$ 5 & 0.60 $\pm$ 0.07 & $-$115.5 $\pm$ 5.8 & 0.26 $\pm$ 0.13 \\ [0.07cm]
&& E1 & 55 $\pm$ 3 & 14.64 $\pm$ 0.60 & $-$114.9 $\pm$ 4.4 & 2.85 $\pm$ 1.20 \\ [0.07cm]
&& E4 & 30 $\pm$ 2 & 10.43 $\pm$ 0.60 & $-$113.9 $\pm$ 7.1 & 1.73 $\pm$ 1.20 \\ [0.07cm]
&& E6 & 60 $\pm$ 3 & 8.38 $\pm$ 0.60 & $-$112.09 $\pm$ 4.4 & 1.43 $\pm$ 1.20 \\ [0.07cm]
&& E8 & 170 $\pm$ 9 & 5.92 $\pm$ 0.26 & $$-$$113.2 $\pm$ 2.3 & 1.03 $\pm$ 0.52 \\ [0.07cm]
&& E9 & 70 $\pm$ 4 & 4.41 $\pm$ 0.50 & $-$112.2 $\pm$ 5.7 & 1.13 $\pm$ 0.99 \\ [0.07cm]
&& S9 & 46 $\pm$ 2 & 1.40 $\pm$ 0.44 & $-$117.9 $\pm$ 9.1 & 0.83 $\pm$ 0.89 \\ [0.07cm]
\hline\\ [0.07cm]
2013.11 & 56333.50 & C0 & 822 $\pm$ 41 &&& 0.11 $\pm$ 0.07 \\ [0.07cm]
&& C1 & 594 $\pm$ 30 & 0.24 $\pm$ 0.01 & $-$117.6 $\pm$ 2.3 & $<$ 0.05 \\ [0.07cm]
&& C2 & 73 $\pm$ 4 & 0.56 $\pm$ 0.01 & $-$114.9 $\pm$ 1.3 & 0.06 $\pm$ 0.05 \\ [0.07cm]
&& E1 & 41 $\pm$ 2 & 14.27 $\pm$ 0.60 & $-$114.5 $\pm$ 6.4 & 2.67 $\pm$ 1.20 \\ [0.07cm]
&& E6 & 77 $\pm$ 4 & 8.61 $\pm$ 0.60 & $-$111.8 $\pm$ 4.6 & 1.61 $\pm$ 1.20 \\ [0.07cm]
&& E8 & 78 $\pm$ 4 & 6.31 $\pm$ 0.20 & $-$112.4 $\pm$ 1.7 & 0.57 $\pm$ 0.41 \\ [0.07cm]
&& E8a & 62 $\pm$ 3 & 5.64 $\pm$ 0.20 & $-$114.6 $\pm$ 1.9 & 0.49 $\pm$ 0.40 \\ [0.07cm]
&& E9 & 87 $\pm$ 4 & 4.52 $\pm$ 0.49 & $-$111.6 $\pm$ 5.1 & 1.25 $\pm$ 0.98 \\ [0.07cm]
&& S9 & 41 $\pm$ 2 & 1.07 $\pm$ 0.21 & $-$117.3 $\pm$ 9.2 & 0.42 $\pm$ 0.42 \\ [0.07cm]
\hline\\ [0.07cm]
2013.16 & 56351.50 & C0 & 922 $\pm$ 46 &&& 0.10 $\pm$ 0.05 \\ [0.07cm]
&& C1 & 642 $\pm$ 32 & 0.24 $\pm$ 0.01 & $-$121.6 $\pm$ 2.1 & $<$ 0.05\\ [0.07cm]
&& C2 & 88 $\pm$ 4 & 0.55 $\pm$ 0.01 & $-$116.5 $\pm$ 1.3 & $<$ 0.05\\ [0.07cm]
&& E1 & 40 $\pm$ 2 & 14.47 $\pm$ 0.60 & $-$114.2 $\pm$ 7.2 & 2.60 $\pm$ 1.20 \\ [0.07cm]
&& E6 & 71 $\pm$ 4 & 8.85 $\pm$ 0.60 & $-$112.6 $\pm$ 4.4 & 1.86 $\pm$ 1.20 \\ [0.07cm]
&& E8 & 84 $\pm$ 4 & 6.44 $\pm$ 0.21 & $-$112.5 $\pm$ 1.7 & 0.61 $\pm$ 0.42 \\ [0.07cm]
&& E8a & 62 $\pm$ 3 & 5.72 $\pm$ 0.20 & $-$115.1 $\pm$ 1.9 & 0.49 $\pm$ 0.39 \\ [0.07cm]
&& E9 & 82 $\pm$ 4 & 4.65 $\pm$ 0.45 & $-$111.9 $\pm$ 6.1 & 1.13 $\pm$ 0.90 \\ [0.07cm]
&& S9 & 53 $\pm$ 3 & 1.12 $\pm$ 0.28 & $-$115.4 $\pm$ 9.5 & 0.60 $\pm$ 0.55 \\ [0.07cm]
\hline\\ [0.07cm]
2013.24 & 56382.50 & C0 & 718 $\pm$ 36 &&& $<$ 0.05 \\ [0.07cm]
&& C1 & 1123 $\pm$ 56 & 0.25 $\pm$ 0.01 & $-$116.9 $\pm$ 2.0 & 0.09 $\pm$ 0.05 \\ [0.07cm]
&& C2 & 114 $\pm$ 6 & 0.60 $\pm$ 0.01 & $-$119.4 $\pm$ 2.9 & 0.07 $\pm$ 0.02 \\ [0.07cm]
&& E1 & 28 $\pm$ 1 & 14.67 $\pm$ 0.60 & $-$113.4 $\pm$ 7.0 & 3.01 $\pm$ 1.20 \\ [0.07cm]
&& E6 & 53 $\pm$ 3 & 9.20 $\pm$ 0.60 & $-$111.9 $\pm$ 5.4 & 1.75 $\pm$ 1.20 \\ [0.07cm]
&& E8 & 75 $\pm$ 4 & 6.76 $\pm$ 0.22 & $-$112.3 $\pm$ 1.7 & 0.59 $\pm$ 0.44 \\ [0.07cm]
&& E8a & 70 $\pm$ 4 & 5.88 $\pm$ 0.20 & $-$114.7 $\pm$ 1.8 & 0.53 $\pm$ 0.39 \\ [0.07cm]
&& E9 & 76 $\pm$ 4 & 4.76 $\pm$ 0.55 & $-$112.2 $\pm$ 5.3 & 1.28 $\pm$ 1.10 \\ [0.07cm]
&& S9 & 44 $\pm$ 2 & 1.43 $\pm$ 0.31 & $-$117.4 $\pm$ 9.5 & 0.61 $\pm$ 0.63 \\ [0.07cm]
\hline\\ [0.07cm]
2013.42 & 56445.50 & C0 & 530 $\pm$ 27 &&& 0.13 $\pm$ 0.06 \\ [0.07cm]
&& C1 & 626 $\pm$ 31 & 0.27 $\pm$ 0.01 & $-$114.9 $\pm$ 2.2 & 0.14 $\pm$ 0.05 \\ [0.07cm]
&& C2 & 369 $\pm$ 18 & 0.63 $\pm$ 0.02 & $-$113.4 $\pm$ 1.5 & 0.15 $\pm$ 0.03 \\ [0.07cm]
&& E1 & 29 $\pm$ 1 & 15.27 $\pm$ 0.60 & $-$114.5 $\pm$ 7.2 & 2.38 $\pm$ 1.20 \\ [0.07cm]
&& E6 & 55 $\pm$ 3 & 9.92 $\pm$ 0.60 & $-$112.7 $\pm$ 5.7 & 2.18 $\pm$ 1.20 \\ [0.07cm]
&& E8 & 67 $\pm$ 3 & 7.16 $\pm$ 0.40 & $-$112.2 $\pm$ 2.9 & 0.92 $\pm$ 0.79 \\ [0.07cm]
&& E8a & 68 $\pm$ 3 & 6.24 $\pm$ 0.15 & $-$114.4 $\pm$ 1.3 & 0.40 $\pm$ 0.30 \\ [0.07cm]
&& E9 & 80 $\pm$ 4 & 5.17 $\pm$ 0.51 & $-$113.3 $\pm$ 5.4 & 1.24 $\pm$ 1.02 \\ [0.07cm]
&& S9 & 64 $\pm$ 3 & 1.37 $\pm$ 0.37 & $-$120.1 $\pm$ 9.4 & 0.84 $\pm$ 0.74 \\ [0.07cm]
\hline\\ [0.07cm]
2013.51 & 56481.50 & C0 & 687 $\pm$ 34 &&& 0.16 $\pm$ 0.06 \\ [0.07cm]
&& C1 & 457 $\pm$ 23 & 0.28 $\pm$ 0.02 & $-$113.5 $\pm$ 3.3 & 0.18 $\pm$ 0.05 \\ [0.07cm]
&& C2 & 235 $\pm$ 12 & 0.79 $\pm$ 0.04 & $-$114.9 $\pm$ 2.9 & 0.27 $\pm$ 0.09 \\ [0.07cm]
&& E1 & 17 $\pm$ 1 & 15.56 $\pm$ 0.60 & $-$116.0 $\pm$ 7.6 & 1.94 $\pm$ 1.20 \\ [0.07cm]
&& E6 & 26 $\pm$ 1 & 10.14 $\pm$ 0.60 & $-$112.2 $\pm$ 6.7 & 1.89 $\pm$ 1.20 \\ [0.07cm]
&& E8 & 46 $\pm$ 3 & 7.46 $\pm$ 0.48 & $-$112.3 $\pm$ 3.4 & 0.89 $\pm$ 0.96 \\ [0.07cm]
&& E8a & 70 $\pm$ 4 & 6.44 $\pm$ 0.18 & $-$113.9 $\pm$ 1.5 & 0.48 $\pm$ 0.35 \\ [0.07cm]
&& E9 & 60 $\pm$ 3 & 5.31 $\pm$ 0.55 & $-$113.5 $\pm$ 5.9 & 1.14 $\pm$ 1.11 \\ [0.07cm]
&& S9 & 40 $\pm$ 2 & 1.50 $\pm$ 0.45 & $-$119.5 $\pm$ 9.4 & 0.79 $\pm$ 0.91 \\ [0.07cm]
\hline\\ [0.07cm]
2013.57 & 56503.50 & C0 & 694 $\pm$ 35 &&& 0.10 $\pm$ 0.07 \\ [0.07cm]
&& C1 & 525 $\pm$ 26 & 0.30 $\pm$ 0.01 & $-$113.6 $\pm$ 2.0 & 0.13 $\pm$ 0.02 \\ [0.07cm]
&& C2 & 139 $\pm$ 7 & 0.72 $\pm$ 0.03 & $-$114.9 $\pm$ 2.4 & 0.16 $\pm$ 0.06 \\ [0.07cm]
&& E1 & 19 $\pm$ 1 & 15.56 $\pm$ 0.60 & $-$115.7 $\pm$ 4.5 & 1.32 $\pm$ 1.20 \\ [0.07cm]
&& E6 & 31 $\pm$ 2 & 10.67 $\pm$ 0.60 & $-$112.0 $\pm$ 8.7 & 1.75 $\pm$ 1.20 \\ [0.07cm]
&& E8 & 47 $\pm$ 2 & 7.70 $\pm$ 0.49 & $-$112.8 $\pm$ 3.3 & 0.91 $\pm$ 0.97 \\ [0.07cm]
&& E8a & 76 $\pm$ 4 & 6.62 $\pm$ 0.18 & $-$113.7 $\pm$ 1.4 & 0.49 $\pm$ 0.35 \\ [0.07cm]
&& E9 & 70 $\pm$ 4 & 5.44 $\pm$ 0.60 & $-$113.9 $\pm$ 5.0 & 1.32 $\pm$ 1.19 \\ [0.07cm]
&& S9 & 115 $\pm$ 6 & 1.07 $\pm$ 0.06 & $-$116.1 $\pm$ 7.0 & 0.26 $\pm$ 0.13 \\ [0.07cm]
\hline\\ [0.07cm]
2013.63 & 56524.50 & C0 & 752 $\pm$ 38 &&& 0.18 $\pm$ 0.07 \\ [0.07cm]
&& C1 & 340 $\pm$ 17 & 0.35 $\pm$ 0.03 & $-$111.4 $\pm$ 4.0 & 0.22 $\pm$ 0.05 \\ [0.07cm]
&& C2 & 158 $\pm$ 8 & 0.98 $\pm$ 0.08 & $-$118.3 $\pm$ 4.6 & 0.38 $\pm$ 0.16 \\ [0.07cm]
&& E1 & 11 $\pm$ 1 & 16.46 $\pm$ 0.58 & $-$119.2 $\pm$ 1.9 & 0.51 $\pm$ 1.15 \\ [0.07cm]
&& E6 & 34 $\pm$ 2 & 11.63 $\pm$ 1.88 & $-$113.4 $\pm$ 6.8 & 2.39 $\pm$ 3.76 \\ [0.07cm]
&& E8 & 31 $\pm$ 2 & 7.89 $\pm$ 0.43 & $-$110.4 $\pm$ 3.0 & 0.66 $\pm$ 0.86 \\ [0.07cm]
&& E8a & 80 $\pm$ 4 & 6.72 $\pm$ 0.21 & $-$112.7 $\pm$ 1.7 & 0.59 $\pm$ 0.42 \\ [0.07cm]
&& E9 & 54 $\pm$ 3 & 5.50 $\pm$ 0.51 & $-$114.1 $\pm$ 5.3 & 1.02 $\pm$ 1.02 \\ [0.07cm]
&& S9 & 19 $\pm$ 1 & 1.71 $\pm$ 0.51 & $-$117.8 $\pm$ 9.0 & 0.60 $\pm$ 1.01 \\ [0.07cm]
\hline\\ [0.1cm]
\label{tab15G}
\end{longtable*}

%____________________________________________________________________________
%____________________________ VLBA-43 GHZ____________________________________
%____________________________________________________________________________

\begin{longtable*}{c c c c c c c}
  \caption[]{- VLBA 43 GHz Model-fit Components' Parameters}
  \endfirsthead
%\hline\\ 
%\multicolumn{7}{|c|}{VLBA 43 GHz} \\ [0.07cm] 
\hline\\ [0.07cm] 
Epoch & Epoch & Name & Flux  & Distance from & Pos. Angle & FWHM  \\ [0.07cm]
(year) & (MJD) & & (mJy) & c0 (mas) & ($^{\circ}$) &  (mas)\\ [0.07cm] 
\hline\\ [0.07cm]
2012.07 & 55953.50 & c0 & 708 $\pm$ 43 & & & 0.08 $\pm$ 0.03\\ [0.07cm] 
&& d7 & 34 $\pm$ 13 & 0.13 $\pm$ 0.01 & $-$124.8 $\pm$ 11.2 & 0.13 $\pm$ 0.04 \\ [0.07cm]
&& d10 & 346 $\pm$ 26 & 0.13 $\pm$ 0.01 & $-$123.7 $\pm$ 8.6 & 0.13 $\pm$ 0.03 \\ [0.07cm]
\hline\\ [0.07cm]
2012.18 & 55991.50 & c0 & 756 $\pm$ 45 & & & 0.08  $\pm$ 0.03 \\ [0.07cm]
&& d7 & 66 $\pm$ 16 & 0.28 $\pm$ 0.05 & $-$123.7 $\pm$ 20.4 & 0.16 $\pm$ 0.03 \\ [0.07cm]
&& d10 & 264 $\pm$ 22 & 0.13 $\pm$ 0.02 & $-$126.2 $\pm$ 11.5 & 0.14 $\pm$ 0.03 \\ [0.07cm]
\hline\\ [0.07cm]
2012.25 & 56019.50 & c0 & 372 $\pm$ 26 & & & 0.06  $\pm$ 0.03\\ [0.07cm]
&& d7 & 42 $\pm$ 13 & 0.41 $\pm$ 0.08 & $-$122.8 $\pm$ 14.2 & 0.18 $\pm$ 0.03 \\ [0.07cm]
&& d10 & 186 $\pm$ 18 & 0.13 $\pm$ 0.01 & $-$113.1 $\pm$ 6.1 & 0.08 $\pm$ 0.03 \\ [0.07cm]
\hline\\ [0.07cm]
2012.40 & 56073.50 & c0 & 680 $\pm$ 41 & & & 0.05  $\pm$ 0.03 \\ [0.07cm]
&& d7 & 33 $\pm$ 11 & 0.46 $\pm$ 0.03 & $-$119.6 $\pm$ 4.4 & 0.07 $\pm$ 0.03\\ [0.07cm]
&& d10 & 195 $\pm$ 17 & 0.18 $\pm$ 0.01 & $-$116.7 $\pm$ 1.6 & 0.04 $\pm$ 0.03\\[0.07cm]
\hline\\ [0.07cm]
2012.51 & 56112.50 & c0 & 667 $\pm$ 40 & & & 0.06  $\pm$ 0.03\\ [0.07cm]
&& d7 & 30 $\pm$ 13 & 0.47 $\pm$ 0.08 & $-$119.2 $\pm$ 11.7 & 0.16 $\pm$ 0.04 \\ [0.07cm]
&& d9 & 66 $\pm$ 13 & 0.26 $\pm$ 0.03 & $-$114.7 $\pm$ 6.1 & 0.09 $\pm$ 0.03 \\ [0.07cm]
&& d10 & 301 $\pm$ 23 & 0.13 $\pm$ 0.01 & $-$112.6 $\pm$ 4.0 & 0.08 $\pm$ 0.03 \\[0.07cm]
\hline\\ [0.07cm]
2012.62 & 56152.50 & c0 & 1044 $\pm$ 59 & & & 0.06 $\pm$ 0.03 \\ [0.07cm]
&& d7 & 52 $\pm$ 16 & 0.66 $\pm$ 0.17 & $-$120.6 $\pm$ 18.5 & 0.40 $\pm$ 0.05 \\ [0.07cm]
&& d8 & 76 $\pm$ 14 & 0.36 $\pm$ 0.04 & $-$112.1 $\pm$ 6.6 & 0.14 $\pm$ 0.05 \\ [0.07cm]
&& d9 & 180 $\pm$ 16 & 0.20 $\pm$ 0.01 & $-$113.9 $\pm$ 1.9 & 0.05 $\pm$ 0.03 \\ [0.07cm]
&& d10 & 503 $\pm$ 33 & 0.09 $\pm$ 0.01 & $-$108.4 $\pm$ 4.0 & 0.08 $\pm$ 0.03 \\ [0.07cm]
\hline\\ [0.07cm]
2012.80 & 56220.50 & c0 & 852 $\pm$ 49 & & & 0.05 $\pm$ 0.03 \\ [0.07cm]
&& d7 & 59 $\pm$ 14 & 0.48 $\pm$ 0.07 & $-$116.5 $\pm$ 9.6 & 0.20 $\pm$ 0.05 \\ [0.07cm]
&& d8 & 136 $\pm$ 16 & 0.35 $\pm$ 0.02 & $-$112.3 $\pm$ 2.8 & 0.09 $\pm$ 0.03 \\ [0.07cm]
&& d9 & 336 $\pm$ 25 & 0.22 $\pm$ 0.01 & $-$111.3 $\pm$ 2.5 & 0.09 $\pm$ 0.03 \\ [0.07cm]
&& d10 & 664 $\pm$ 40 & 0.08 $\pm$ 0.01 & $-$117.3 $\pm$ 4.1 & 0.08 $\pm$ 0.03 \\ [0.07cm]
\hline\\ [0.07cm]
2012.83 & 56229.50 & c0 & 848 $\pm$ 49 & & & 0.06 $\pm$ 0.03 \\ [0.07cm]
&& d7 & 66 $\pm$ 15 & 0.51 $\pm$ 0.08 & $-$116.4 $\pm$ 9.9 & 0.23 $\pm$ 0.03 \\ [0.07cm]
&& d8 & 98 $\pm$ 13 & 0.36 $\pm$ 0.01 & $-$110.8 $\pm$ 1.5 & 0.05 $\pm$ 0.03 \\ [0.07cm]
&& d9 & 329 $\pm$ 25 & 0.23 $\pm$ 0.01 & $-$113.9 $\pm$ 2.9 & 0.10 $\pm$ 0.03 \\ [0.07cm]
&& d10 & 531 $\pm$ 34 & 0.09 $\pm$ 0.01 & $-$115.6 $\pm$ 3.8 & 0.07 $\pm$ 0.03 \\ [0.07cm]
\hline\\ [0.07cm]
2012.97 & 56283.50 & c0 & 1143 $\pm$ 64 & & & 0.07 $\pm$ 0.03 \\ [0.07cm]
&& d8 & 53 $\pm$ 3 & 0.44 $\pm$ 0.01 & $-$121.2 $\pm$ 1.0 & $<$0.03 \\ [0.07cm]
&& d9 & 344 $\pm$ 26 & 0.27 $\pm$ 0.01 & $-$112.7 $\pm$ 2.6 & 0.11 $\pm$ 0.03 \\ [0.07cm]
&& d10 & 422 $\pm$ 29 & 0.10 $\pm$ 0.01 & $-$116.3 $\pm$ 5.5 & 0.09 $\pm$ 0.03\\ [0.07cm]
\hline\\ [0.07cm]
2013.04 & 56307.50 & c0 & 1227 $\pm$ 67 & & & 0.03 $\pm$ 0.03  \\ [0.07cm]
&& d8 & 153 $\pm$ 16 & 0.36 $\pm$ 0.01 & $-$114.3 $\pm$ 2.5 & 0.08 $\pm$ 0.03 \\ [0.07cm]
&& d9 & 328 $\pm$ 23 & 0.21 $\pm$ 0.01 & $-$112.9 $\pm$ 0.9 & 0.03 $\pm$ 0.03 \\ [0.07cm]
&& d10 & 882 $\pm$ 50 & 0.07 $\pm$ 0.01 & $-$95.5 $\pm$ 0.8 & 0.03 $\pm$ 0.03 \\ [0.07cm]
\hline\\ [0.07cm]
2013.15 & 56349.50 & c0 & 879 $\pm$ 51 & & & 0.06 $\pm$ 0.03 \\ [0.07cm]
&& d8 & 143 $\pm$ 16 & 0.41 $\pm$ 0.02 & $-$114.8 $\pm$ 2.6 & 0.09 $\pm$ 0.03 \\ [0.07cm]
&& d9 & 467 $\pm$ 24 & 0.23 $\pm$ 0.01 & $-$110.1 $\pm$ 1.0 &  0.03 $\pm$ 0.03 \\ [0.07cm]
&& d10 & 828 $\pm$ 49 & 0.14 $\pm$ 0.01 & $-$103.4 $\pm$ 1.7 & 0.08 $\pm$ 0.03 \\ [0.07cm]
\hline\\ [0.07cm]
2013.29 & 56398.50 & c0 & 441 $\pm$ 29 & & & 0.06 $\pm$ 0.03 \\ [0.07cm]
&& d8 & 464 $\pm$ 31 & 0.45 $\pm$ 0.01 & $-$114.1 $\pm$ 1.4 & 0.11 $\pm$ 0.03 \\ [0.07cm]
&& d9 & 408 $\pm$ 29 & 0.25 $\pm$ 0.01 & $-$113.2 $\pm$ 3.1 & 0.12 $\pm$ 0.03 \\ [0.07cm]
&& d10 & 617 $\pm$ 38 & 0.09 $\pm$ 0.01 & $-$119.5 $\pm$ 3.6 & 0.07 $\pm$ 0.03 \\ [0.07cm]
\hline\\ [0.07cm]
2013.41 & 56442.50 & c0 & 586 $\pm$ 36 & & & 0.07 $\pm$ 0.03\\ [0.07cm]
&& d9 & 340 $\pm$ 26 & 0.34 $\pm$ 0.02 & $-$111.1 $\pm$ 4.1 & 0.20 $\pm$ 0.03 \\ [0.07cm]
&& d10 & 441 $\pm$ 30 & 0.09 $\pm$ 0.01 & $-$99.2 $\pm$ 4.3 & 0.09 $\pm$ 0.03 \\ [0.07cm]
&& d11 & 211 $\pm$ 21 & 0.70 $\pm$ 0.04 & $-$112.8 $\pm$ 2.7 & 0.22 $\pm$ 0.03 \\ [0.07cm]
\hline\\ [0.07cm]
2013.49 & 56473.50 & c0 & 639 $\pm$ 39 & & & 0.07 $\pm$ 0.03\\ [0.07cm]
&& d8 & 138 $\pm$ 17 & 0.46 $\pm$ 0.03 & $-$112.7 $\pm$ 4.4 & 0.16 $\pm$ 0.03 \\ [0.07cm]
&& d9 & 191 $\pm$ 18 & 0.28 $\pm$ 0.01 & $-$106.6 $\pm$ 2.5 & 0.09 $\pm$ 0.03 \\ [0.07cm]
&& d10 & 523 $\pm$ 34 & 0.11 $\pm$ 0.01 & $-$102.8 $\pm$ 2.6 & 0.08 $\pm$ 0.03 \\ [0.07cm]
&& d11 & 123 $\pm$ 17 & 0.90 $\pm$ 0.05 & $-$112.9 $\pm$ 2.7 & 0.21 $\pm$ 0.03 \\ [0.07cm]
\hline\\ [0.07cm]
2013.57 & 56501.50 & c0 & 567 $\pm$ 35 & & & 0.06 $\pm$ 0.03 \\ [0.07cm]
&& d8 & 194 $\pm$ 19 & 0.46 $\pm$ 0.3 & $-$111.7 $\pm$ 3.6 & 0.17 $\pm$ 0.03 \\ [0.07cm]
&& d9 & 210 $\pm$ 18 & 0.24 $\pm$ 0.01 & $-$104.6 $\pm$ 1.8 & 0.06 $\pm$ 0.03\\ [0.07cm]
&& d10 & 380 $\pm$ 27 & 0.13 $\pm$ 0.01 & $-$104.9 $\pm$ 2.7 & 0.07 $\pm$ 0.03 \\ [0.07cm]
&& d11 & 90 $\pm$ 15 & 1.06 $\pm$ 0.05 & $-$113.6 $\pm$ 2.3 & 0.18 $\pm$ 0.03 \\ [0.07cm]
\hline\\ [0.07cm]
2013.65 & 56530.50 & c0 & 567 $\pm$ 36 & & & 0.09 $\pm$ 0.03 \\ [0.07cm]
&& d9 & 180 $\pm$ 19 & 0.33 $\pm$ 0.03 & $-$107.9 $\pm$ 5.2 & 0.18 $\pm$ 0.03 \\ [0.07cm]
&& d10 & 611 $\pm$ 32 & 0.12 $\pm$ 0.01 & $-$102.4 $\pm$ 3.2 & 0.10 $\pm$ 0.03 \\ [0.07cm]
&& d11 & 58 $\pm$ 14 & 1.20 $\pm$ 0.07 & $-$114.6 $\pm$ 3.1 & 0.20 $\pm$ 0.03 \\ [0.07cm]
\hline\\ [0.07cm]
2013.88 & 56614.50 & c0 & 788 $\pm$ 46 & & & 0.05 $\pm$ 0.03 \\ [0.07cm]
&& d9 & 367 $\pm$ 26 & 0.26 $\pm$ 0.01 & $-$104.9 $\pm$ 1.3 & 0.07 $\pm$ 0.03 \\ [0.07cm]
&& d10 & 677 $\pm$ 41 & 0.12 $\pm$ 0.01 & $-$105.3 $\pm$ 1.6 & 0.06 $\pm$ 0.03 \\ [0.07cm]
&& d11 & 51 $\pm$ 15 & 1.69 $\pm$ 0.10 & $-$116.8 $\pm$ 3.3 & 0.26 $\pm$ 0.04 \\ [0.07cm]
&& d11a & 12 $\pm$ 1 & 2.07 $\pm$ 0.14 & $-$114.9 $\pm$ 3.1 & 0.17 $\pm$ 0.03 \\ [0.07cm]
&& d12 & 346 $\pm$ 26 & 0.45 $\pm$ 0.01 & $-$109.5 $\pm$ 1.6 & 0.13 $\pm$ 0.03 \\ [0.07cm]
\hline\\ [0.07cm]
2013.96 & 56642.50 & c0 & 623 $\pm$ 39 & & & 0.09 $\pm$ 0.03 \\ [0.07cm]
&& d10 & 409 $\pm$ 29 & 0.16 $\pm$ 0.01 & $-$113.9 $\pm$ 5.7 & 0.14 $\pm$ 0.03 \\ [0.07cm]
&& d11 & 58 $\pm$ 15 & 1.72 $\pm$ 0.13 & $-$116.6 $\pm$ 4.0 & 0.32 $\pm$ 0.03 \\ [0.07cm]
&& d11a & 36 $\pm$ 14 & 2.22 $\pm$ 0.12 & $-$115.6 $\pm$ 3.7 & 0.25 $\pm$ 0.04 \\ [0.07cm]
&& d12 & 223 $\pm$ 21 & 0.59 $\pm$ 0.03 & $-$111.1 $\pm$ 2.6 & 0.19 $\pm$ 0.03 \\ [0.07cm]
&& d12a & 202 $\pm$ 19 & 0.40 $\pm$ 0.02 & $-$109.6 $\pm$ 2.5 & 0.13 $\pm$ 0.03 \\ [0.07cm]
\hline\\ [0.07cm]
2014.05 & 56677.50 & c0 &  639 $\pm$ 39 & & & 0.07 $\pm$ 0.03 \\ [0.07cm]
&& d9 & 98 $\pm$ 14 & 0.27 $\pm$ 0.03 & $-$106.2 $\pm$ 5.2 & 0.11 $\pm$ 0.03 \\ [0.07cm]
&& d10 & 322 $\pm$ 24 & 0.13 $\pm$ 0.01 & $-$103.3 $\pm$ 4.1 & 0.09 $\pm$ 0.03 \\ [0.07cm]
&& d11 & 54 $\pm$ 15 & 1.86 $\pm$ 0.15 & $-$115.0 $\pm$ 3.7 & 0.36 $\pm$ 0.03 \\ [0.07cm]
&& d11a & 32 $\pm$ 14 & 2.37 $\pm$ 0.15 & $-$116.0 $\pm$ 4.4 & 0.29 $\pm$ 0.05 \\ [0.07cm]
&& d12 & 119 $\pm$ 17 & 0.51 $\pm$ 0.06 & $-$108.5 $\pm$ 2.8 & 0.22 $\pm$ 0.03 \\ [0.07cm]
&& d12a & 79 $\pm$ 15 & 0.51 $\pm$ 0.06 & $-$108.5 $\pm$ 4.2 & 0.22 $\pm$ 0.03 \\ [0.07cm]
\hline\\ [0.07cm]
2014.15 & 56713.50 & c0 & 468 $\pm$ 30 & & & 0.06 $\pm$ 0.03  \\ [0.07cm]
&& d8 & 95 $\pm$ 14 & 0.39 $\pm$ 0.03 & $-$108.6 $\pm$ 4.1 & 0.12 $\pm$ 0.03 \\ [0.07cm]
&& d9 & 109 $\pm$ 15 & 0.23 $\pm$ 0.03 & $-$105.2 $\pm$ 5.7 & 0.12 $\pm$ 0.03 \\ [0.07cm]
&& d10 & 338 $\pm$ 25 & 0.08 $\pm$ 0.01 & $-$107.9 $\pm$ 4.9 & 0.07 $\pm$ 0.03\\ [0.07cm]
&& d11 & 57 $\pm$ 16 & 2.09 $\pm$ 0.19 & $-$114.5 $\pm$ 3.3 & 0.46 $\pm$ 0.06 \\ [0.07cm]
&& d11a & 8 $\pm$ 2 & 2 $\pm$ 0.12 & $-$115.1 $\pm$ 3.1 & 0.12 $\pm$ 0.03 \\ [0.07cm]
&& d12 & 47 $\pm$ 14 & 0.65 $\pm$ 0.13 & $-$111.8 $\pm$ 2.9 & 0.20 $\pm$ 0.03 \\ [0.07cm]
&& d12a & 20 $\pm$ 13 & 0.65 $\pm$ 0.13 & $-$111.8 $\pm$ 4.2 & 0.20 $\pm$ 0.03 \\ [0.07cm]
\hline\\ [0.07cm]
2014.33 & 56780.50 & c0 & 633 $\pm$ 38 & & & 0.04 $\pm$ 0.03 \\ [0.07cm]
&& d8 & 76 $\pm$ 14 & 0.40 $\pm$ 0.03 & $-$110.5 $\pm$ 4.7 & 0.12 $\pm$ 0.03 \\ [0.07cm]
&& d9 & 146 $\pm$ 16 & 0.24 $\pm$ 0.02 & $-$113.5 $\pm$ 4.3 & 0.09 $\pm$ 0.03 \\ [0.07cm]
&& d10 & 463 $\pm$ 30 & 0.10 $\pm$ 0.01 & $-$113.7 $\pm$ 2.3 & 0.05 $\pm$ 0.03 \\ [0.07cm]
&& d11 & 48 $\pm$ 15 & 2.47 $\pm$ 0.12 & $-$114.2 $\pm$ 2.6 & 0.29 $\pm$ 0.04 \\ [0.07cm]
&& d12 & 29 $\pm$ 13 & 1.49 $\pm$ 0.09 & $-$111.4 $\pm$ 2.8 & 0.18 $\pm$ 0.04 \\ [0.07cm]
&& d12a & 54 $\pm$ 14 & 0.87 $\pm$ 0.08 & $-$110.5 $\pm$ 4.8 & 0.22 $\pm$ 0.04 \\ [0.07cm]
\hline\\ [0.1cm]
  \label{tab43G}
\end{longtable*}

\end{document}